\documentclass[aps,prb,twocolumn,superscriptaddress,longbibliography]{revtex4-2}

\usepackage[colorlinks=true,citecolor=blue]{hyperref}
\usepackage{graphicx}
\usepackage{amsmath}
\usepackage{amssymb}

\DeclareGraphicsExtensions{.png,.jpg,.eps}
\usepackage{xcolor}

\DeclareMathOperator{\EX}{\mathbb{E}}

\newcommand{\comment}[1]{}

\begin{document}

\title{Designing quantum many-body matter with conditional generative adversarial networks
}
\author{Rouven Koch}
\affiliation{Department of Applied Physics, Aalto University, 00076 Aalto, Espoo, Finland}
\author{Jose L. Lado}
\affiliation{Department of Applied Physics, Aalto University, 00076 Aalto, Espoo, Finland}

\date{\today}

\begin{abstract}
The computation of dynamical correlators of quantum many-body systems represents an open critical challenge in condensed matter physics. While powerful methodologies have risen in recent years, covering the full parameter space remains unfeasible for most many-body systems with a complex configuration space. Here we demonstrate that conditional Generative Adversarial Networks (GANs) allow simulating the full parameter space of several many-body systems, accounting both for controlled parameters, and stochastic disorder effects. 
After training with a restricted set of noisy many-body calculations, the conditional GAN algorithm provides the whole dynamical excitation spectra for a Hamiltonian instantly and with an accuracy analogous to the exact calculation. We further demonstrate how the trained conditional GAN automatically provides a powerful method for Hamiltonian learning from its dynamical excitations, and to flag non-physical systems via outlier detection. Our methodology puts forward generative adversarial learning as a powerful technique to explore complex many-body phenomena, providing a starting point to design large-scale quantum many-body matter. 
\end{abstract}

\maketitle

\section{Introduction}

The dynamical properties of quantum many-body models remain one of the critical problems in condensed matter physics, lying at the heart of problems ranging from correlated superconductivity~\cite{RevModPhys.66.763} to quantum spin liquid physics~\cite{Savary2016,RevModPhys.88.041002}. 
Even with the appearance of powerful new methodologies in the last years~\cite{PhysRevLett.69.2863,Carleo2017}, tackling specific regimes of quantum many-body models is an outstanding problem~\cite{Zheng2017,PhysRevX.5.041041} and covering the full parameter space of a many-body Hamiltonian quickly is a nearly unfeasible task. 
This huge complexity is not a feature alone of quantum many-body physics, but it is also well known in many problems of image, voice, and video recognition~\cite{bin2017highquality,Gao2018,2017arXiv170907592X}. In these fields, a new family of algorithms known as \textit{Generative Adversarial Networks}~(GANs)~\cite{goodfellow2014generative} has allowed to tackle some of those intractable problems with high accuracy~\cite{isola2017image,8629024,2020arXiv200812073N}. 

While supervised and unsupervised learning has been widely applied to quantum problems~\cite{Carrasquilla2017,Carrasquilla2020,RevModPhys.91.045002,PhysRevX.7.031038,vanNieuwenburg2017,PhysRevLett.123.230504,PhysRevB.99.245120,Greplova2020,PhysRevB.98.060301,PhysRevB.97.115453,2017arXiv171105238G,RodriguezNieva2019,PhysRevLett.124.226401}, generative adversarial learning remains relatively unexplored~\cite{kenig2021tunable,liu2017simulating,ahmed2021quantum}.
The advantages of GANs over simple (supervised or unsupervised) \textit{neural network}~(NN) models are the ability of learning underlying distributions of complex data set (e.g., images) and the generation of new samples with the same statistics by only using input noise (and additional conditional parameters)~\cite{radford2016unsupervised,perarnau2016invertible}. The generated output is of such high accuracy, e.g., photo-realistic images, that can not be achieved similarly with other generative models~\cite{2016arXiv160603498S}. Moreover, GANs naturally incorporate noise in the generative network architecture which enables to account for both uncertainty and diversity in the model. This includes \textit{multi-modal learning} where one input can correspond to several correct outputs which can not be achieved by classical machine learning algorithms which generally learn a one-to-one mapping~\cite{2020arXiv200106937G}.

\begin{figure}[t]
    \includegraphics[width=.98\columnwidth]{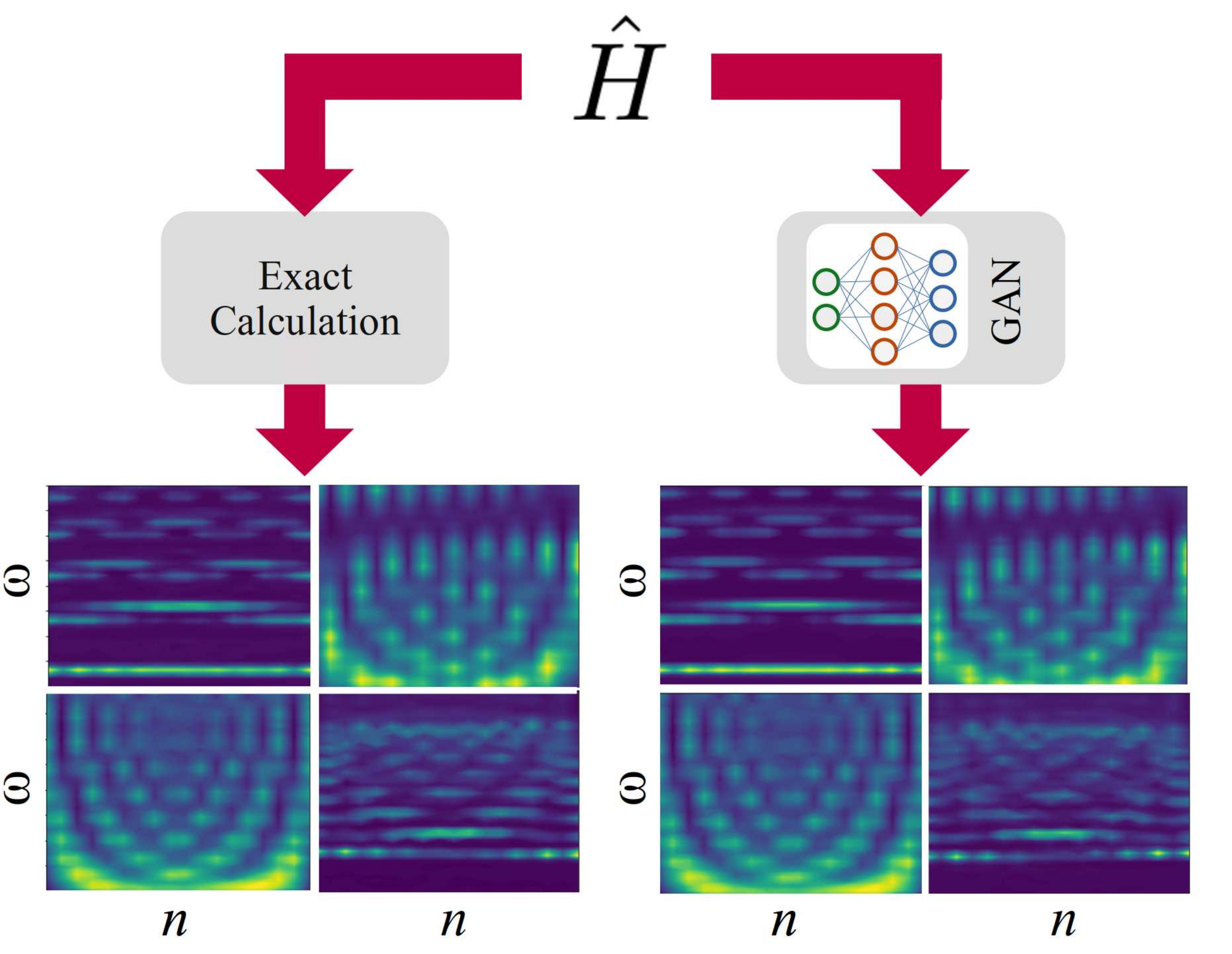}
    \caption{Comparison of the dynamical correlator computed with an exact many-body formalism (left) and a trained conditional generative adversarial neural network (right). After training, the cGAN allows generating many-body spin correlators of analogous quality to the many-body formalism in the whole parameter space. The trained cGAN accounts simultaneously both for controlled Hamiltonian parameters and hidden disorder effects.}
    \label{fig:intro}
\end{figure}

Here we show how \textit{conditional} GANs (cGANs) allow simulating dynamical excitations of many-body Hamiltonians and furthermore provide efficient Hamiltonian learning and outlier detection.
Taking as training examples a finite set of noisy many-body dynamical calculations, we demonstrate that the conditional GAN quickly learns to generate dynamical results for the whole parameter space (as illustrated in Fig.~\ref{fig:intro}). Once the GANs are trained, the computational and generalization power of GANs over traditional methods comes into play: even to simulate new many-body Hamiltonians of big system size, the outputs of the GAN are almost instantaneous and with an accuracy rivaling the exact calculations, enabling a detailed mapping of complex many-body systems without the need to calculate every parameter combination. 
Besides realizing a powerful simulator, the trained GAN automatically provides two additional features by exploiting the trained discriminator. First, the parameters of the Hamiltonian can be directly inferred from the simulated dynamical data by using the cGAN discriminator, a methodology providing a cGAN-based Hamiltonian learning algorithm.
Second, the trained discriminator allows detecting non-physical results such as those stemming from wrongly computed dynamical many-body systems. Our work provides a first step towards designing quantum many-body matter with deep generative models, opening a pathway to address complex quantum many-body landscapes and ultimately combining theoretical and experimental data.

The manuscript is organized as follows. Sec.~\ref{sec:GAN} introduces the general concept of cGANs and the quantum many-body methodology for computing dynamical correlators with tensor networks. As a first demonstration, Sec~\ref{sec:single} exemplifies our cGAN methodology for a family of single-particle models. Sec.~\ref{sec:many} demonstrates the cGAN methodology for three families of quantum many-body systems, including a gapless many-body model featuring spinons, a model with topological order, and a fermionic Hubbard model. In Sec.\ref{sec:extra}, we are showing the extrapolation capability of our algorithm and giving a quantitative benchmark of the cGAN.
Section~\ref{sec:disc} demonstrates how the trained cGAN provides both a methodology for Hamiltonian learning and outlier detection. Finally, Sec.~\ref{sec:summary} summarizes our conclusions. Information about the GAN architecture and training data generation are given in App.~\ref{app:A}, and App.~\ref{app:B} and in App.~\ref{app:C} we are providing a supplementary analysis of the generator and discriminator network.

\section{Generative Adversarial Networks and dynamical correlators} 
\label{sec:GAN}

\subsection{Generative Adversarial Networks} 

Generative Adversarial Networks were proposed in 2014 as deep generative models in the context of unsupervised Machine Learning (ML)~\cite{goodfellow2014generative}. They are generally built by combining two neural networks, the generator \textit{G} and discriminator \textit{D}, which are competing in a min-max game against each other. This allows the generator to become very accurate in mapping from a latent space vector \textbf{z}~(i.e., a random input vector) to the data distribution of the real images. The generator network tries to trick the discriminator which has the job of distinguishing between real and generated images. During the training process, the parameters of both networks get updated simultaneously, minimizing the terms related to the generator $\text{log}\left[1-\text{D}(\text{G}(z))\right]$ and discriminator $\text{log}[\text{D}(x)]$ which are part of the GAN value function
\begin{equation}
    \label{Eq:val1}
    \begin{split}
        \text{min}_\text{G}\, \text{max}_\text{D} V(\text{D},\text{G}) & = \EX_{x\thicksim p_{\text{data}}(\textbf{x})} \left[\text{log} \text{D}(\textbf{x})\right] \\[1mm]
        & + \EX_{z\thicksim p_z(\textbf{z})} [\text{log}~(1-\text{D}(\text{G}(\textbf{z}))] \, .
    \end{split}
\end{equation}
 The input data \textbf{x} contains the information of the real images, $p_{\text{data}}$ is the distribution of the input images which we want to learn, and $p_{z}$ is the (normal) distribution of the latent space. During the training, the parameters of the generator (discriminator) are updated in order to minimize (maximize) the expectation values of the value function $V(\text{D},\text{G})$.

\begin{figure}[t]
    \includegraphics[width=1\columnwidth]{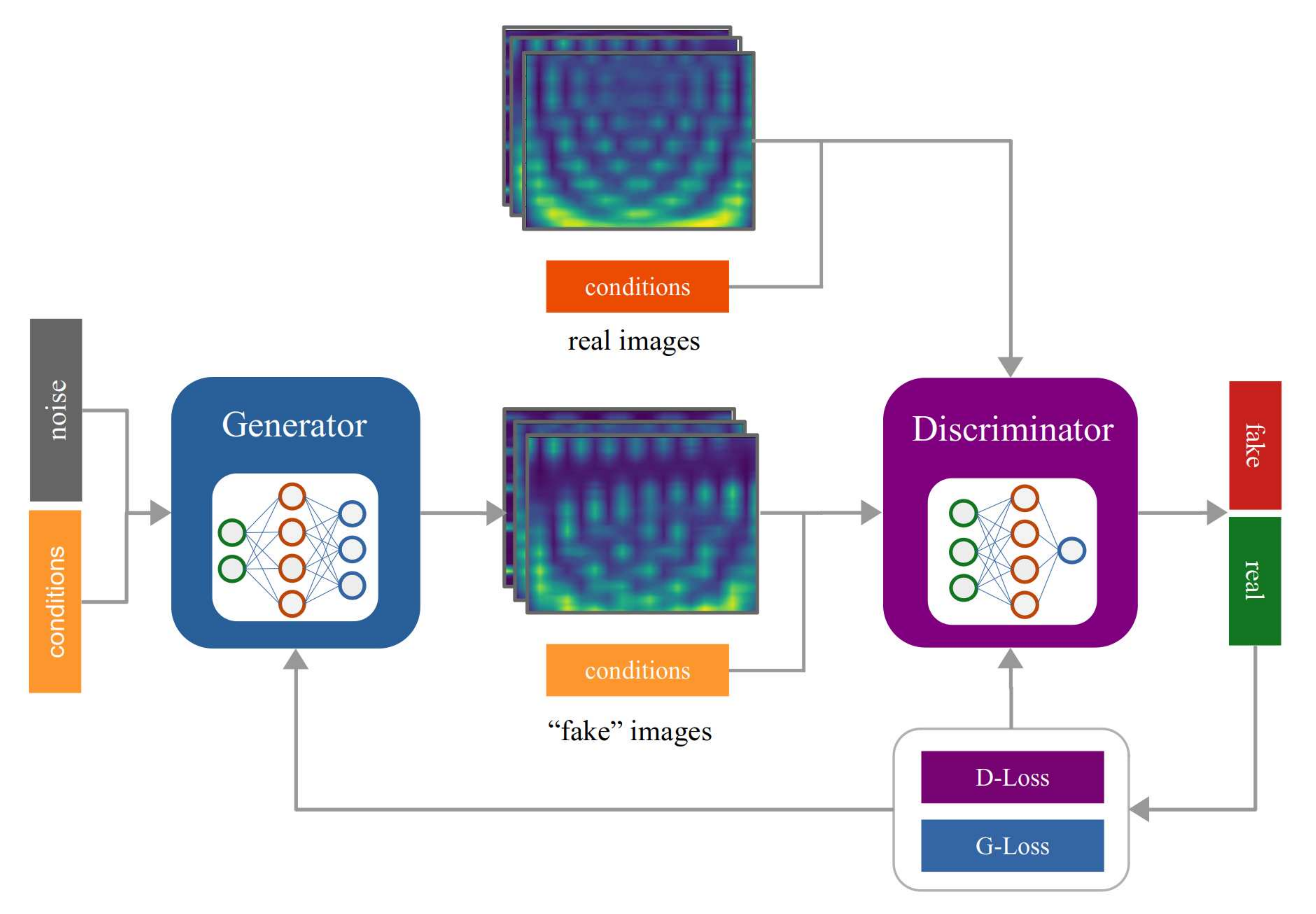}
    \caption{Illustration of the cGAN architecture used in this work. The cGAN consists of two deep neural networks, the generator G and discriminator D, playing a min-max game against each other. The generator learns the (labeled) data distribution from the real images and the discriminator tries to distinguish between real and generated images. The network parameters are updated during the training via the D-Loss and G-Loss.
    }
        \label{fig:GAN}
\end{figure}

Applications of GANs in computer science tasks usually include the generation of images including convolutional neural networks and have shown great success in the applications of image generation by using random inputs~\cite{radford2016unsupervised,karras2018progressive,brock2019large}. These random inputs, however, inhibit us from controlling the output of the algorithm. An extension to the usual GAN are cGANs which give additional information to the neural networks in order to gain some control over the output of the algorithm without losing the generative power of this method~\cite{mirza2014conditional}. Some applications include, e.g,. image-to-image translation~\cite{isola2017image} or image editing~\cite{perarnau2016invertible, bin2017highquality, antipov2017face, zhang2019image}. The computational power of (conditional) GANs has already found its way to physics, starting in high energy physics for the simulation of 2D particle jet images~\cite{de2017learning} and 3D particle showers~\cite{paganini2018accelerating}, cosmology for emulations of cosmological maps~\cite{perraudin2021emulation}, and in selected problems of quantum and condensed matter physics including the simulation of correlated Quantum Walk~\cite{kenig2021tunable} and to simulate 2D Ising model near the critical temperature~\cite{liu2017simulating}. Recently, conditional GANs have also been successfully applied for quantum state tomography and the reconstruction of density matrices~\cite{ahmed2021quantum}.

In particular, conditional GANs allow for the incorporation of prior knowledge about a system and simultaneously account for a degree of diverse randomness in the output. This architecture corresponds to cGANs which have a vector of labels (\textbf{y}) in addition to the training data as input of the generator and discriminator. In the specific case of our manuscript, we consider conditional labels that are given by the different energy scales of a general Hamiltonian. Figure~\ref{fig:GAN} shows the general architecture of the cGAN used in this work. This architecture is inspired by conventional GANs, yet with the key difference that conditional parameters are included as input for the generator and discriminator (shown in orange). The value function is also very similar to the one of Eq.~\ref{Eq:val1}~\cite{mirza2014conditional}

\begin{equation}
    \label{Eq:val2}
    \begin{split}
        \text{min}_\text{G}\, \text{max}_\text{D} V(\text{D},\text{G}) & = \EX_{x\thicksim p_{\text{data}}(\textbf{x})} [\text{log} \text{D}(\textbf{x}|\textbf{y})] \\[1mm]
        & + \EX_{z\thicksim p_z(\textbf{z})} [\text{log}~(1-\text{D}(\text{G}(\textbf{z}|\textbf{y}))]
    \end{split}
\end{equation}

with conditional constraints \textbf{y} in the input of the discriminator and generator in their corresponding term in the value function. In the case of image generation, the auxiliary labels of the cGAN have discrete class values. In our case, we are using continuous labels which allows us to cover the full parameter space of a given Hamiltonian with a continuous cGAN. In contrast to conventional GANs, we now have the ability to simulate many-body systems with conditional parameters in the Hamiltonian. 

\subsection{Dynamical correlators with tensor-networks} 
\label{sec:dynamics}

Here we summarize the many-body method used to generate the training data. We will be interested in computing the dynamical correlator of a many-body Hamiltonian, taking the form

\begin{equation}
\label{eq:correlator}
\chi(\omega)  =
        \langle GS| \hat A \delta (\omega \mathcal{I} - \hat H+E_{GS}) \hat B | GS \rangle 
\end{equation}
where $\hat A,\hat B, \hat H$ are many-body operators and $| GS \rangle $ is the many-body ground state
and $E_{GS}$ is the ground state energy. 
This spectral function corresponds to the dynamical spin structure factor for a spin system and the electronic many-body density of states for an electronic system. 

We now elaborate on the dynamical correlator $S_z(\omega,n)$, that corresponds to the local spin structure factor~\cite{Giamarchi2003,Mahan2000,RevModPhys.91.041001}. From the physical point of view, the dynamical spin structure factor signals the existence of spin excitations at a specific energy in a material~\cite{Giamarchi2003,Mahan2000,RevModPhys.91.041001}. From the experimental point of view, spin excitations such excitations can be directly measured via inelastic spectroscopy with scanning tunnel microscope~\cite{PhysRevLett.102.256802,PhysRevLett.111.127203,Loth2010,Toskovic2016,Spinelli2014}.
The spin excitations are directly probed by tunneling electrons, in which an electron with spin up tunnels to the magnetic system, flipping its spin, creating a spin excitation, and tunnels outside of it~\cite{PhysRevLett.102.256802}.
This process gives rise to a step in the differential conductance $dI/dV$~\cite{PhysRevLett.102.256802}, and in turn, directly appears as a peak in the $d^2 I/dV^2$~\cite{Spinelli2014}. The spin excitations computed in our manuscript are therefore directly measured experimentally, as demonstrated in a variety of experiments with scanning tunneling microscope~\cite{Loth2010,Toskovic2016,Spinelli2014,RevModPhys.91.041001}.

The dynamical correlator is computed using the tensor-network kernel polynomial algorithm~\cite{RevModPhys.78.275,PhysRevB.90.115124,PhysRevResearch.2.023347,PhysRevB.90.045144,PhysRevResearch.3.013095,PhysRevResearch.1.033009,PhysRevResearch.3.013265,PhysRevResearch.3.033102}. The many-body states and Hamiltonians are represented in terms of a tensor-network, using the matrix-product state formalism~\cite{2020arXiv200714822F,ITensor,dmrgpy}, the ground state is computed with the density-matrix renormalization group algorithm~\cite{PhysRevLett.69.2863}, and the Hamiltonian is scaled to the interval $(-1,1)$ to perform the Chebyshev expansion\cite{RevModPhys.78.275}. The scaled Hamiltonian is denoted as $\bar H$, and its scaled spectral function as $\bar \chi$, taking the form
 \begin{equation}
     \bar \chi(x) = \frac{1}{\pi\sqrt{1-x^2}}
     \left [ \alpha_0 + 2 \sum_{n=1}^\infty \alpha_n T_n(x) \right ]
 \end{equation}
 with $T_n(x)$ the Chebyshev polynomials and $\alpha_n$ the coefficients of the expansion computed recursively, and including the Jackson Kernel~\cite{Jackson1912}. Finally, we note that while we focus here on the tensor-network representation of the states, an analogous procedure can be performed with neural-network quantum states~\cite{PhysRevB.104.205130}.
 
\newcommand{\chemp} {\mu}
\newcommand{\simb} {m}

\section{Single-particle systems}
\label{sec:single}
\begin{figure}[t]
    \includegraphics[width=.99\columnwidth]{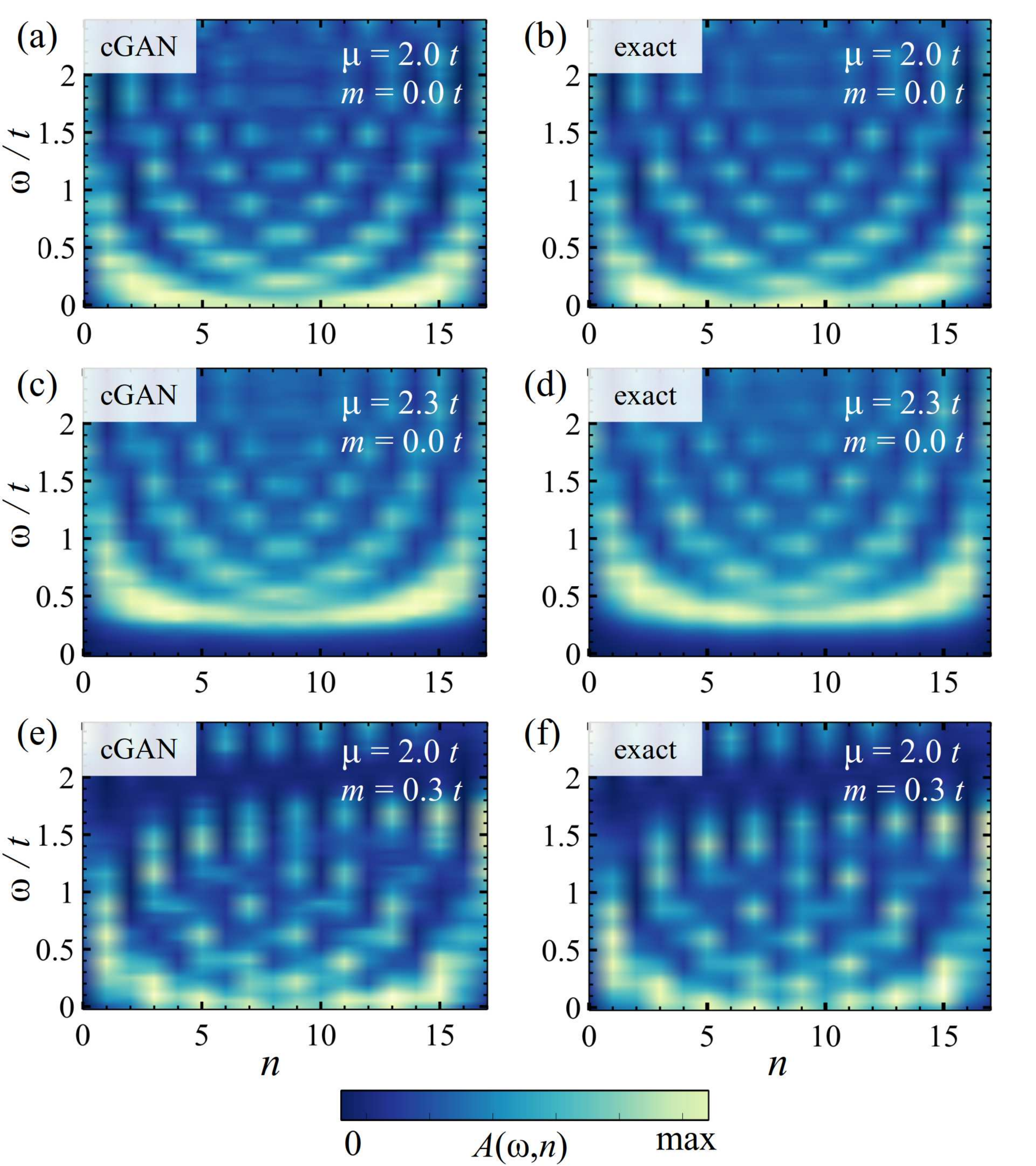}
    \caption{Real space DOS of a one-dimensional tight-binding system of length $n=18$ generated with the cGAN (a,c,e) in comparison with the exact tight-binding calculations (b,d,f). The x-axis labels the site, the y-axis the frequency, and the z-axis the (local) DOS (Eq.~\ref{Eq:ldos}). Shown are 3 combinations of the conditional parameters ($\chemp , \simb$) in (a,b), (c,d) and (e,f), as defined in Eq.~\ref{Eq:tb}.
    }
    \label{fig:results_tb}
\end{figure}

While ultimately we will explore our generative algorithm for a quantum many-body system,
it is instructive to first explore its applicability for a family of single-particle models
that can easily be solved. As the first proof of concept, we test our cGAN for a one-dimensional single-particle tight-binding system. The Hamiltonian in second quantization of these systems is given by

\begin{equation}
    \label{Eq:tb}
    \begin{split}
        H(\chemp, \simb) &= t \sum_{n} c_n^\dagger c_{n+1} +\text{h.c.} + \chemp \sum_{n} \,c_n^\dagger c_n \\ 
        & \hspace{1.7mm} + \simb \sum_{n}  (-1)^{n} c_n^\dagger c_n   + \sum_n  v_n c_n^\dagger c_n \, ,
    \end{split}
\end{equation}
with the hopping $t$, $v_{n}$ random onsite energies, and $\mu$ as chemical potential. The additional fourth term in the equation introduces a site imbalance with magnitude $\simb$ and defines together with the chemical potential the conditional parameter space of our Hamiltonian. The onsite energies $v_{n}$ are chosen randomly in the interval $v_{n} \in [-0.1, 0.1]\,t$ to introduce the randomness in the training data of the cGAN. 
This randomness emulates potential hidden variables in the model, small fluctuations associated with the theoretical methodology, and would allow mimicking additional perturbations which could be present in potential future experimental data and were not accounted by for the theoretical model.
We computed 4000 real systems and extended the training set with the data-enhancement method presented in App.\ref{app:B} to 32 000 examples.\footnote{This enhancement procedure is not critical for a simple single-particle system but becomes significantly more important for the computational costly many-body systems studied in the next section.} This training set size is therefore in the order of the MNIST data set of handwritten digits \cite{deng2012mnist}. The parameters $\chemp$ and $\simb$ are the conditional parameter of the GAN and are defined in the intervals $\chemp \in [1.7, 2.3]\,t$ and $\simb \in [-0.3, 0.3]\,t$.

The idea is to train the generator to map from ($\chemp , \simb$) to the (local) density of states $A(\omega, n)$ (DOS) which is defined as   
\begin{equation}
\label{Eq:ldos}
     A(\omega, n) =  
\langle n | \delta( \omega - H) | n \rangle 
\end{equation}
where $\mathcal{H}$ is the tight binding matrix defined by Eq.~\ref{Eq:tb}, and $\delta$ the Dirac delta function.

The density of states $A(\omega,n)$ corresponds to the spectra of charge excitations of the system~\cite{Mahan2000}. In particular, it directly corresponds to the probability of an electron with specific energy to tunnel to a specific location~\cite{PhysRevLett.50.1998,Binnig1983}.
A non-zero density of states at certain frequency signals that a single electron would be able to tunnel into the material at such energy~\cite{PhysRevLett.50.1998}. From the experimental point of view, the electron spectral function can be directly probed via scanning tunneling spectroscopy~\cite{Binnig1983,PhysRevB.31.805,RevModPhys.59.615}. In particular, the differential conductance defined as $dI/dV$ allows to directly access the electron spectral function of a material~\cite{PhysRevLett.50.1998,RevModPhys.59.615}, directly corresponding to the quantity computed in our manuscript.
The density of states has been directly measured in a variety of setups, and in particular directly allows probing the spatial distribution of quantized modes~\cite{Crommie1993,Heller1994,RevModPhys.75.933}.

Figure~\ref{fig:results_tb} shows the value of the DOS (z-axis) depending on the site (x-axis) and frequency (y-axis). We show the spatial-resolved DOS for 3 different conditional parameter combinations ($\chemp , \simb$) and compare the simulations of the cGAN in Fig.~\ref{fig:results_tb}~(a,c,e) with the exact calculations in Fig.~\ref{fig:results_tb}~(b,d,f). 
As observed in the figure, there is no visual difference between the real and generated DOS for each of the three parameter choices, a feature observed for generic examples. In particular, in Fig.~\ref{fig:results_tb}~(c,d), the increase of $\chemp$ gives rise to a frequency shift of $0.3\,t$ compared to Fig.~\ref{fig:results_tb}~(a,b), which is very well captured by the generated DOS of the cGAN in (c). Similar results can be seen in Fig.~\ref{fig:results_tb}~(e,d), where the increased $\simb$-parameter induces a site imbalance between odd and even sites in the chain. In conclusion, the simulations of the algorithm capture the effects of both conditional parameters on the DOS with high accuracy and in arbitrary magnitude. The trained generator is able to generate new systems with arbitrary parameter choice of ($\chemp , \simb$) in the boundaries of the training interval, and even slightly outside, almost instantaneous with very high precision. In the next section, the same algorithm is applied to three different many-body systems which are computational more demanding than the single-particle system which can be seen as proof of principle.

\section{Many-body systems}
\label{sec:many}
\begin{figure}[t]
    \includegraphics[width=.99\columnwidth]{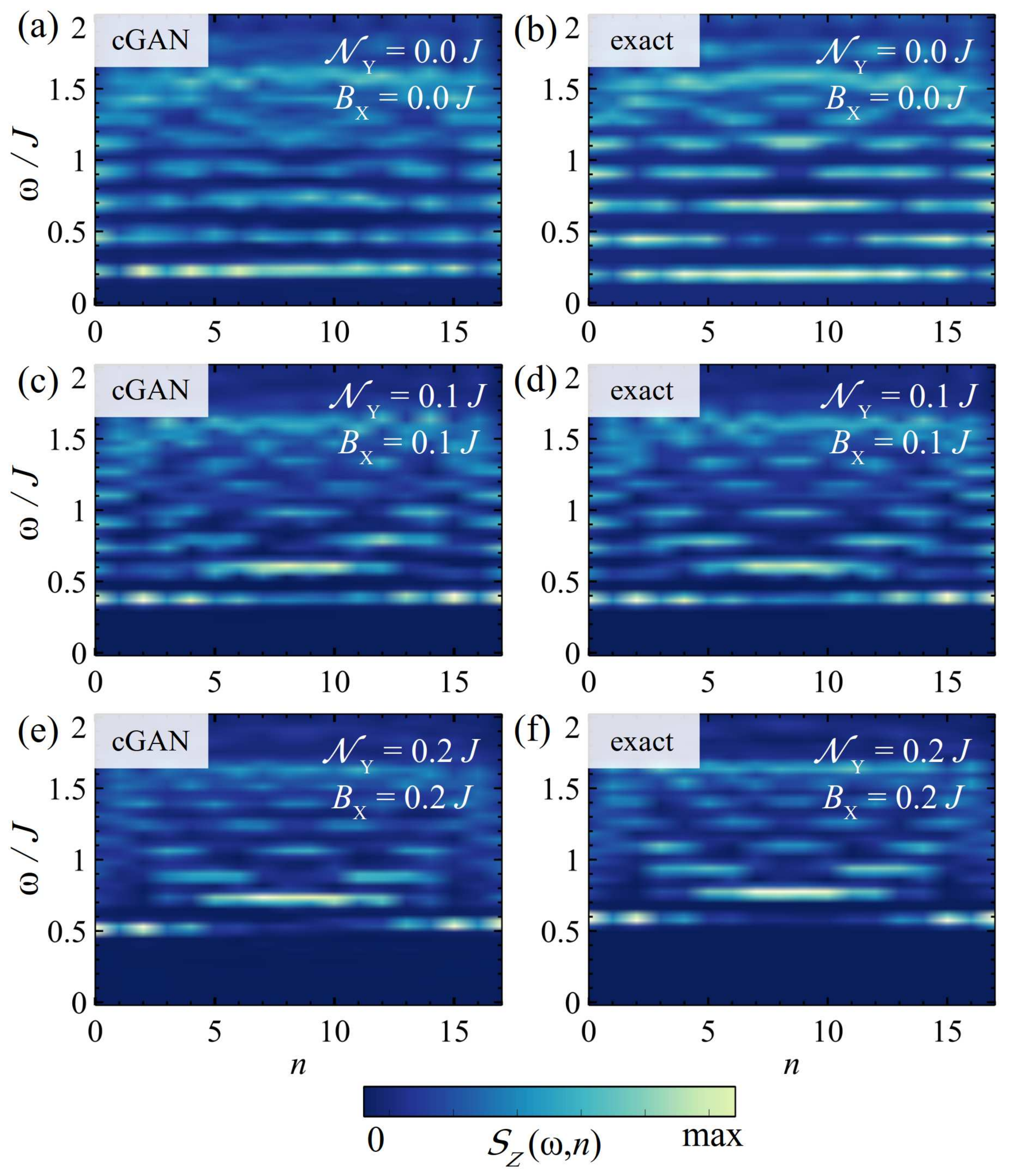}
    \caption{
    Real space $S^z$ spin correlators of a one-dimensional $S=1/2$-spin chain of length $n=18$ generated with the cGAN (a,c,e) in comparison with exact tensor-network calculations (b,d,f). The x-axis labels the site, the y-axis the frequency, and the z-axis the full spin correlator (Eq.~\ref{eq:dcz}). Shown are 3 combinations of the conditional parameters ($\mathcal{N}_y , B_x$) in (a,b), (c,d), and (e,f), defined in Eq.~\ref{Eq:spinchain}.}
    \label{fig:results_S12}
\end{figure}

In contrast to the single-particle case in the previous section, calculations of many-body systems with high accuracy are computationally much more demanding. This affects the training of the cGAN because creating an arbitrary large training set becomes one of the major bottlenecks. The idea is to use the minimal amount of data to train the network accurately and use methods of data enhancements to enlarge the training set (see App.~\ref{app:B}). This minimizes the computational effort and takes full advantage of the generative power of this algorithm. In this section, we test our cGAN algorithm for three different one-dimensional many-body systems including a $S=1/2$-chain, a topologically non-trivial $S=1$ system, and a doped Hubbard model~\cite{codecgan}. For each Hamiltonian, we have chosen specific conditional parameters and added hidden parameters that, e.g., account for residual perturbations in an experimental setup.

\subsection{Gapless many-body $S=1/2$ spin model}
We start with the simplest many-body system we studied, an interacting
$S=1/2$ Heisenberg model realizing a quantum-disordered ground state.
The Hamiltonian for the one-dimensional $S=1/2$ system is given by 
\begin{equation}
    \label{Eq:spinchain}
    \begin{split}
        H(\mathcal{N}_y, B_x) & = J \, \sum_{n}  {\bold{S}}_n \cdot {\bold{S}}_{n+1} + 
          \mathcal{N}_y \sum_n  (-1)^n S_n^{x}\\
          & \hspace{-0.5mm}+ B_x \sum_n S_n^{y} + \sum_n \left( \xi^x_n  S_n^{x}  + \xi^y_n S_{n}^{y} \right)   
    \end{split}
\end{equation}

with $\bold{S}_n = (S^x_n,S^y_n,S^z_n)$ the $S=1/2$ many-body spin operators.
The parameter $J$ denotes the Heisenberg exchange coupling, $\mathcal{N}_y$ a local alternating Neel magnetic field in the $y$-direction, and $B_x$ a uniform Zeeman field in the $x-$direction. In the absence of Neel, Zeeman, and disorder fields, this model realizes a well-understood isotropic Heisenberg model. In this limit, the system features gapless $S=1/2$ spinon excitations~\cite{Faddeev1984} hosting a spin-singlet ground state with local zero magnetization in the thermodynamic limit~\cite{Giamarchi2003} and can be analytically solved via Bethe ansatz~\cite{Bethe1931}. In the presence of finite Neel and Zeeman terms, the ground state of the system develops a finite order in the $x$ and $y$ directions $\langle S^x_n \rangle \ne 0, \langle S^y_n \rangle \ne 0$, yet hosting a zero local order in the $z$-direction $\langle S^z_n \rangle = 0$.
For our cGAN, the parameters $\mathcal{N}_y$ and $B_x$ are the conditional parameters, defined in the intervals $\mathcal{N}_y \in [0.0, 0.2]\,J$ and $B_x \in [0.0, 0.2]\,J$, and $\xi^x$ and $\xi^y$ are introducing randomness (up to $0.05\,J$) to the training data generation. 

We now focus on the spin excitations in real space computed with the dynamical spin correlator defined as 
\newcommand{\DC}{\mathcal{S}}
\begin{equation}
\DC_z(\omega,n) = \langle GS |\hat S^z_n \delta (\omega \mathcal{I} - \hat H + E_{GS}) \hat S^z_n |GS \rangle
\label{eq:dcz}
\end{equation}
where $\hat S^z_n$ is the local spin operator in site $n$, $|GS \rangle$ the many-body ground state, and $E_{GS}$ the ground state energy. The previous correlator directly probes many-body spin excitations in the spin chain and can be directly measured experimentally in real space~\cite{RevModPhys.91.041001} using inelastic spectroscopy~\cite{Heinrich2004,Spinelli2014,Toskovic2016} and electrically-driven paramagnetic resonance with scanning tunneling microscopy~\cite{Baumann2015,PhysRevLett.119.227206,Willke2019,Yang2019,Yang2021}. We train the cGAN to map from the conditional parameters to the correlator in real space ($\mathcal{N}_y , B_x) \rightarrow \DC$. For the training we used 2250 many-body calculations with arbitrary conditional parameter combinations and used data-enhancement methods (shown in App.~\ref{app:B}) to increase the training set size to 36 000.

The results for the $S=1/2$ system for 3 different parameter combinations are shown in Fig.~\ref{fig:results_S12}~(a,b), Fig.~\ref{fig:results_S12}~(c,d) and Fig.~\ref{fig:results_S12}~(e,f). We compare the simulated systems in Fig.~\ref{fig:results_S12}~(a,c,e) with the real many-body calculations in Fig.~\ref{fig:results_S12}~(d,e,f). The parameter combinations cover different areas of the parameter space of the Hamiltonian of Eq.~\ref{Eq:spinchain} and are chosen randomly. The cGAN simulates the spin excitations in $z$-direction with high accuracy and captures the important features including the spatial profile of the many-body modes in the full frequency range. Differences for the 3 parameter combinations occur in the form of a shift of the lowest excitation and the location and number of higher many-body modes. In Fig.~\ref{fig:results_S12}~(a,b) the lowest excitation is at $0.25\,J$ which is captured well in the generated system in (a). Especially for $\mathcal{N}_y=0.1\,J$ and $B_x=0.1\,J$ the simulation (Fig.~\ref{fig:results_S12}~(c)) is very close to the real spectrum (Fig.~\ref{fig:results_S12}~(d)) comparing the energy onset of the excitation at around $0.5t$ as well as the relative magnitudes of higher many-body excitations. The same applies to the third parameter combination of $\mathcal{N}_y=0.2\,J$ and $B_x=0.2\,J$ in Fig.~\ref{fig:results_S12}~(e) and Fig.~\ref{fig:results_S12}~(f), respectively. Differences between the simulations and real images can be related to, first, the induced noise up to $0.05\,J$ and the random sampling of the cGAN from this noise distribution, i.e., every system generated by the cGAN shows small but observable differences. The second source of error is connected with the small amount of real training data which implies that for each arbitrary combination of conditional parameters only a small number of training examples exists (in the vicinity of the parameter space). Despite these features, the results for these values and arbitrary parameter combinations are very precise considering the comparatively small amount of training data in terms of GANs (we used only 2250 real data points compared the e.g. the MNIST data set of 60 000 examples) and the instantaneous generation of the spectra.

\subsection{Interacting many-body system with topological order}
The one-dimensional $S=1$ spin chain is a topological non-trivial system that shows spin fractionalization in form of excitations of $S=1/2$ spins below the bulk gap on the edges of the chain~\cite{PhysRevLett.59.799,PhysRevLett.50.1153,PhysRevB.81.064439,PhysRevB.80.155131,PhysRevB.85.075125}. This model represents one of the simplest examples of many-body fractionalization stemming from topological order. This system shows robust topological edge modes, resilient to perturbations that do not break the spin rotational symmetry of the model~\cite{PhysRevB.81.064439,PhysRevB.83.035107}, and has been realized both in natural compounds~\cite{PhysRevB.54.R6827} and artificial designer platforms~\cite{Mishra2021}. In stark contrast with the model of the previous section, the dynamical spectra of this topological model show persistent edge excitations, together with bulk modes, providing a substantially different qualitative behavior.

The Hamiltonian of the spin $S=1$ Heisenberg model we consider is given by  
\begin{equation}
    \label{Eq:spinchain_S1}
    \begin{split}
        H(\Delta_J, J_2) = & \, J \sum_{n} {\bold{S}}_n \cdot {\bold{S}}_{n+1} + J_2 \sum_n  {\bold{S}}_n \cdot {\bold{S}}_{n+2}  \\
        + &  \Delta_J \sum_{n}~(-1)^n \, {\bold{S}}_n \cdot {\bold{S}}_{n+1} \\
        + & \sum_n \,\xi^J_n \, {\bold{S}}_n \cdot {\bold{S}}_{n+1} 
        + \sum_n \, \xi^{J_2}_n  \,{\bold{S}}_n \cdot {\bold{S}}_{n+2}
    \end{split}
\end{equation}
with $\bold{S}_n = ({S}^x_n, {S}^y_n, {S}^z_n) $ the many-body spin operators for $S=1$.
In comparison to Eq.~\ref{Eq:spinchain}, we have now chosen the dimerization of the nearest-neighbor exchange ($\Delta_J$) and second-nearest-neighbor exchange ($J_2$) as conditional parameters. We note that external magnetic fields would break the protection of the low-energy topological excitations of the fractionalized spins of this system which we want to study, and, therefore, are not included. In turn, we introduce two noise terms in the model which respect the topological class, in particular spatially-dependent fluctuation in the exchange $\xi^J_n$ and second-neighbor exchange $\xi^{J_2}_n$. Those two random fluctuations would account for small spatial inhomogeneities of the system in an experimental realization~\cite{PhysRevB.54.R6827,Mishra2021} stemming from local defects. The conditional parameters are defined in the intervals $J_2 \in [-0.2, 0.2]\,J$ and $\Delta_J \in [-0.2, 0.2]\,J$, and $\xi^J_n$ and $\xi^{J_2}_n$ are introducing randomness up to $0.05\,J$. In this case, the cGAN learns a mapping from ($J_2, \Delta_J$) to the full spin correlator

\begin{equation}
\DC(\omega,n) = \sum_\alpha \langle GS |\hat S^\alpha_n \delta (\omega \mathcal{I} - \hat H + E_{GS}) \hat S^\alpha_n |GS \rangle
\label{eq:dcfull}
\end{equation}

that denotes spin excitations in real space. 
It is worth noting that, due to the spin isotropy of Eq.~\ref{Eq:spinchain_S1}, the correlator of Eq.~\ref{eq:dcfull} is proportional to Eq.~\ref{eq:dcz} for the considered $S=1$ model, and can be measured analogously in engineered spin chains with scanning tunneling probes~\cite{RevModPhys.91.041001,Heinrich2004,Spinelli2014,Toskovic2016,Baumann2015,PhysRevLett.119.227206,Willke2019,Yang2019,Yang2021}.

\begin{figure}[t!]
    \includegraphics[width=.99\columnwidth]{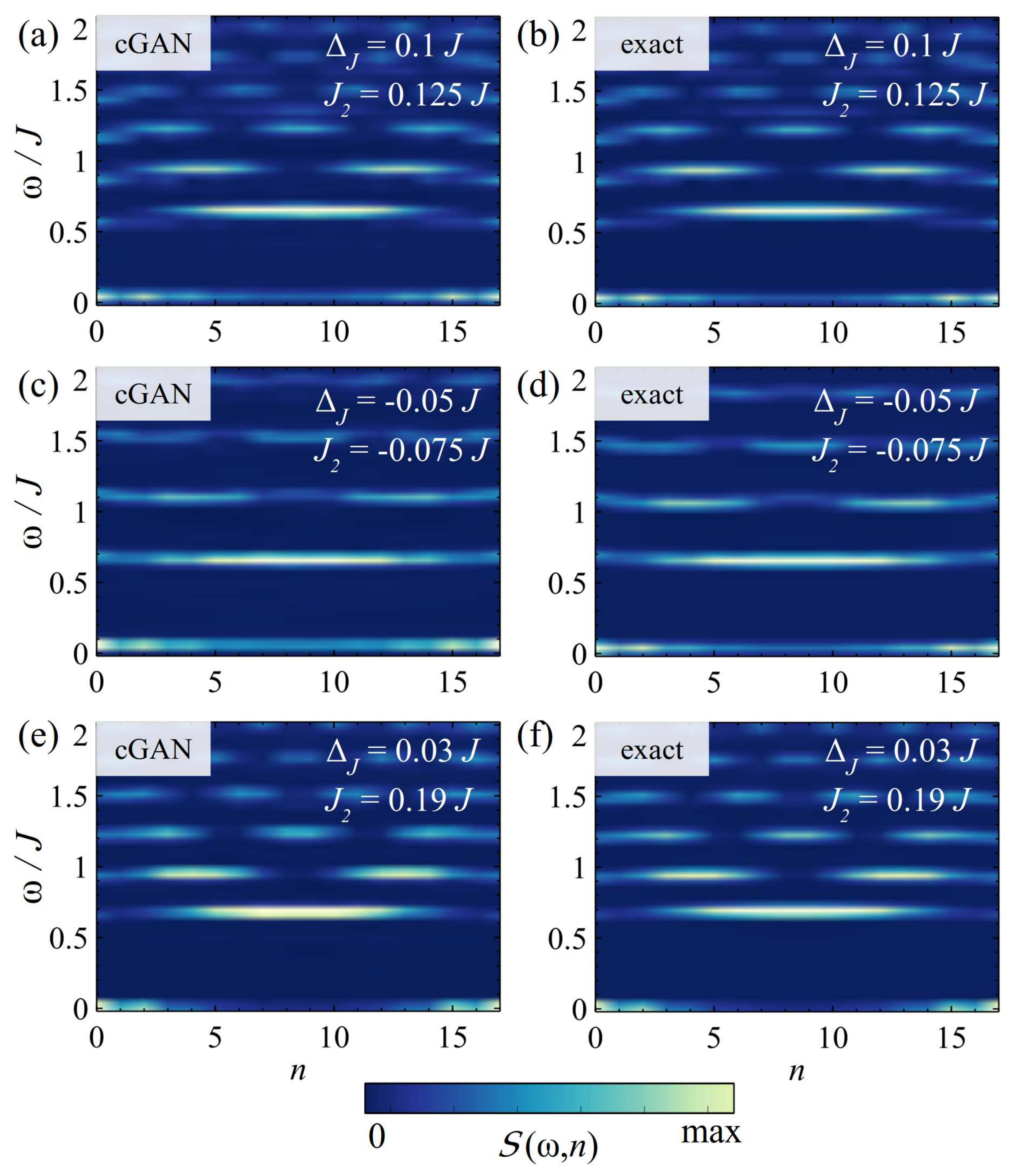}
    \caption{
    Real space $S^z$ spin correlators of a one-dimensional $S=1$-spin chain of length $n=18$ generated with the cGAN (a,c,e) in comparison with exact tensor-network calculations (b,d,f). The x-axis labels the site, the y-axis the frequency, and the z-axis the full spin correlator (Eq.~\ref{eq:dcfull}). Shown are 3 combinations of the conditional parameters ($\Delta_J , J_2$) in (a,b), (c,d), and (e,f), defined in Eq.~\ref{Eq:spinchain_S1}.
    }
        \label{fig:results_S1}
\end{figure}

In Fig.~\ref{fig:results_S1} we have chosen three arbitrary parameter combinations in order to compare the real spin excitations in Fig.~\ref{fig:results_S1}~(b,d,f) with the simulations of the cGAN in Fig.~\ref{fig:results_S1}~(a,c,e).
For all parameter combinations, the fractionalized $S=1/2$ excitations emerge at the edges of the chain close to $\omega=0\,J$ and are mostly not affected by variation of the first and second-neighbor interactions. Due to finite size effects, the fractionalized excitations have a non-zero magnitude even in the middle of the chain for all parameter combinations, due to the dependence of the topological gap on the parameters $J_2$ and $\Delta_J$. This effect is, however, of different magnitude for different choices of $J_2$ and $\Delta_J$. In case of $\Delta_J=0.03\,J$ and $J_2=0.19\,J$ (Fig.~\ref{fig:results_S1}~(e) and (f)), the $S=1/2$ excitations appear mostly close to the edges at site $n=0$ and $n=17$. This behavior is captured well by the generated system in Fig.~\ref{fig:results_S1}~(e).
A stronger first-neighbor dimerization as well as second neighbor exchange interaction (Fig.~\ref{fig:results_S1}~(a) and Fig.~\ref{fig:results_S1}~(b)) results in closer lying excitations above the bulk gap at around $0.7\,J$. This effect is very accurately captured by the cGAN predictions in Fig.~\ref{fig:results_S1}~(e)(b). The values of the energy levels, as well as relative magnitudes, are predicted with high accuracy in comparison to the exact tensor-network calculations for all arbitrarily chosen parameter combinations in the defined intervals. The visual accuracy obtained for this model even surpasses the case of the $S=1/2$ system. This can be related to the spectra themselves which show more pronounced and separated features. To summarize this section, the cGAN is able to simulate the $S=1$ system with high accuracy almost immediately in the range of the introduced randomness under consideration of the minimal amount of training data, same as in $S=1/2$.

\subsection{Interacting fermionic systems}

\begin{figure}[t!]
    \includegraphics[width=.99\columnwidth]{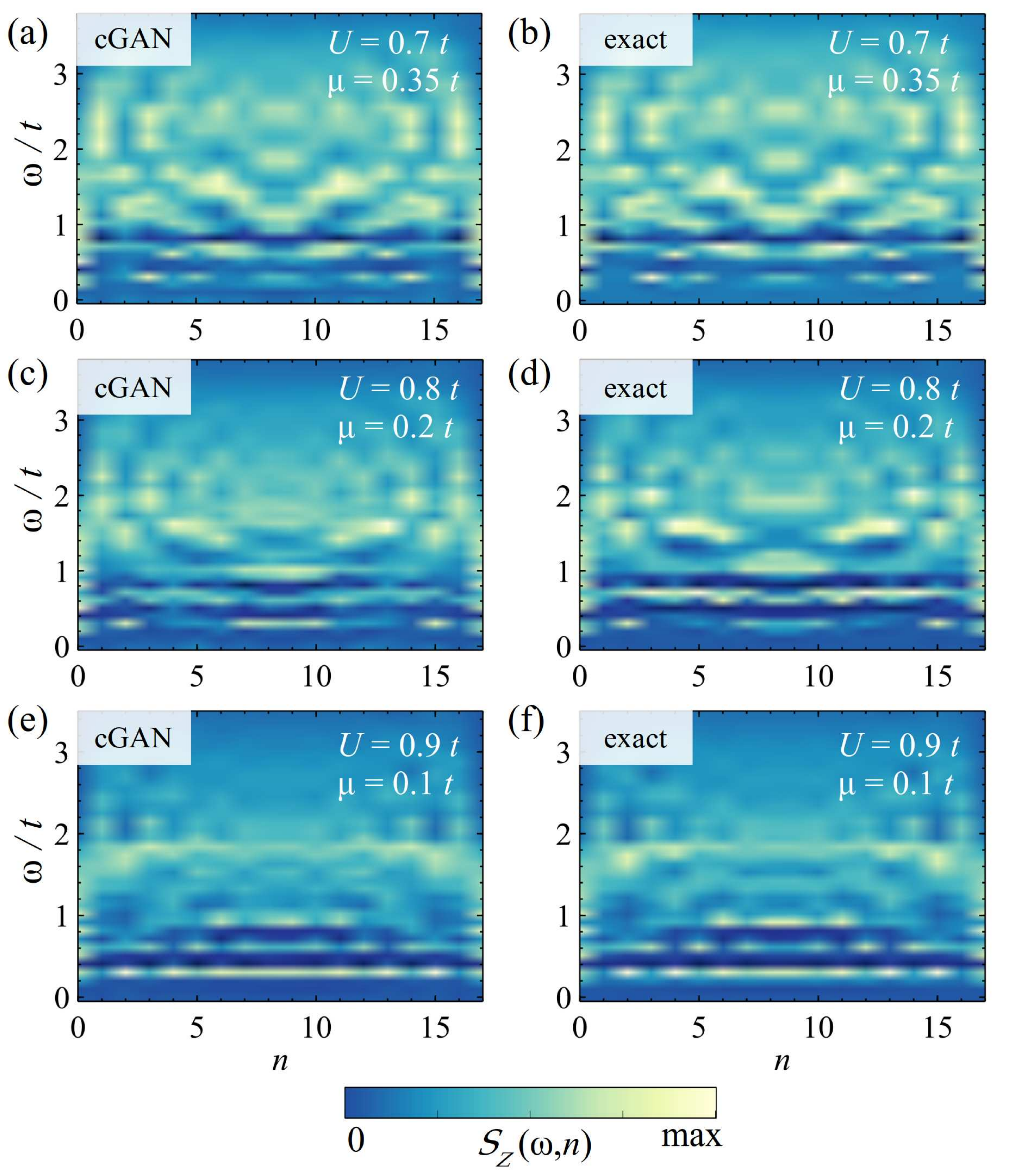}
    \caption{
     Real space $S^z$ spin correlators of a one-dimensional Hubbard model of length $n=18$ generated with the cGAN (a,c,e) in comparison with exact tensor-network calculations (b,d,f). The x-axis labels the site, the y-axis the frequency, and the z-axis the spin correlator in z-direction (Eq.~\ref{eq:dczf}). Shown are 3 combinations of the conditional parameters ($U , \mu$) in (a,b), (c,d), and (e,f), as defined in Eq.~\ref{Eq:hubbard}.
    }
        \label{fig:results_hubbard}
\end{figure}

We now move on to an interacting model with richer many-body phenomena, in particular incorporating both charge and spin degrees of freedom. The third system we are studying with the cGAN is the doped Hubbard model described by the Hamiltonian

\begin{equation}
    \label{Eq:hubbard}
    \begin{split}
        H(U, \mu) & =  t \sum_{n,s} \hspace{1mm} c^\dagger_{n,s} c_{n+1,s} + \mu \sum_{s} c^\dagger_{n,s} c_{n, s} \\
        & + U \sum_n \, (\rho_{n, \uparrow} - 1/2) (\rho_{n, \downarrow} - 1/2)\\
        & + \sum_n \xi_n (-1)^n S^y_{n} \, ,
    \end{split}
\end{equation}

where $\rho_{n, s} = c^\dagger_{n,s} c_{n,s}$, $S^y_{n} = \sum_{s,s'} \sigma^y_{s,s'} c^\dagger_{n,s} c_{n,s'} $,
with $\sigma^y_{s,s'}$ the spin Pauli matrix.
The previous Hamiltonian is well known to feature a widely rich phase diagram away from half-filling~\cite{Essler2005} and provides a paradigmatic example of spin-charge separation~\cite{Voit1995}. In particular, for $\mu=0$ and $U\gg t$ the electronic system is half-filled, and the spin sector of Eq.~\ref{Eq:hubbard} maps to a Heisenberg model with an exchange coupling given by $J \sim t^2/U$ for local $S=1/2$ degrees of freedom.
That limit corresponds to the model presented in Eq.~\ref{Eq:spinchain}. However, in the general case away from half-filling, the previous model shows much more complex spin excitation than Eq.~\ref{Eq:hubbard}. 
The conditional parameters are the onsite Hubbard interaction $U$, chosen in the interval $ U \in [0.5,1]\,t$, and $\mu \in [0.0,0.5]\,t$ that parametrizes the chemical potential of the system. The randomness is created with an external stagger magnetic field in $y$-direction with parameter $\xi$ that alternates sign between neighboring sites ($\xi_n \in [0.0,0.05]\,t$). 
We will focus on addressing the many-body spin excitations of this interacting fermionic model, as given by the dynamical spin correlator

\begin{equation}
\begin{split}
\DC_z(\omega,n) & = \langle GS |S^z_n \delta (\omega - H + E_{GS}) S^z_n |GS \rangle \\ 
& = \frac{1}{4}\langle GS | \left ( \rho_{n,\uparrow} - \rho_{n,\downarrow} \right ) \delta (H - \omega + E_{GS}) \\
& \hspace{8mm} \left ( \rho_{n,\uparrow} - \rho_{n,\downarrow} \right ) |GS \rangle
\end{split}
\label{eq:dczf}
\end{equation}
now written with fermionic many-body operators $\rho_{n,s} = c^\dagger_{n,s} c_{n,s}$.

The cGAN learns the mapping ($U , \mu) \rightarrow \DC_z(\omega,n)$ and the results are presented in Fig.~\ref{fig:results_hubbard} showing the many-body excitations $\DC_z(\omega,n)$ on the corresponding site (x-axis) in the frequency $\omega$ range between 0 and $5\,t$ (y-axis). For the three different combinations of the conditional parameters ($U , \mu$) shown in Fig.~\ref{fig:results_hubbard}~(a,b), Fig.~\ref{fig:results_hubbard}~(c,d), and Fig.~\ref{fig:results_hubbard}~(e,f), the generated spectra of the cGAN Fig.~\ref{fig:results_hubbard}~(a,c,e) show very good agreement with the tensor-network calculations Fig.~\ref{fig:results_hubbard}~(b,d,f). The variation of the onsite interaction $U$ between $0.7\,t$ and $0.9\,t$ as well as the charge occupation varying between $0.1\,t$ and $0.35\,t$ does not affect the accuracy of the generated spin excitations. The features and changes of the corresponding parameter combinations, including energy gap as well as location and intensity of states, are all well captured in the simulations of the cGAN.

Considering the results for the three studied many-body systems, we observe that cGANs are able to capture dynamical correlators of one-dimensional systems with high precision. This methodology can easily be extended to different systems without further modifications of the network architecture. The almost instantaneous simulations provide a huge advantage over the numerically costly tensor-network calculations and enable to study the full parameter space of a Hamiltonian without additional computational effort. Despite the relatively small amount of training data (about one magnitude less than for conventional training of GANs as mentioned earlier in this section) the accuracy remains high and the benefits of the cGAN algorithm out-weight the computational costs of creating the training data. As seen in these examples, the cGAN algorithm provides faithful results for substantially different many-body systems, therefore suggesting that this methodology can be readily extended to other many-body systems. 

\section{Extrapolation and quantitative benchmark of the cGAN}
\label{sec:extra}

\begin{figure}[h]
    \includegraphics[width=.99\columnwidth]{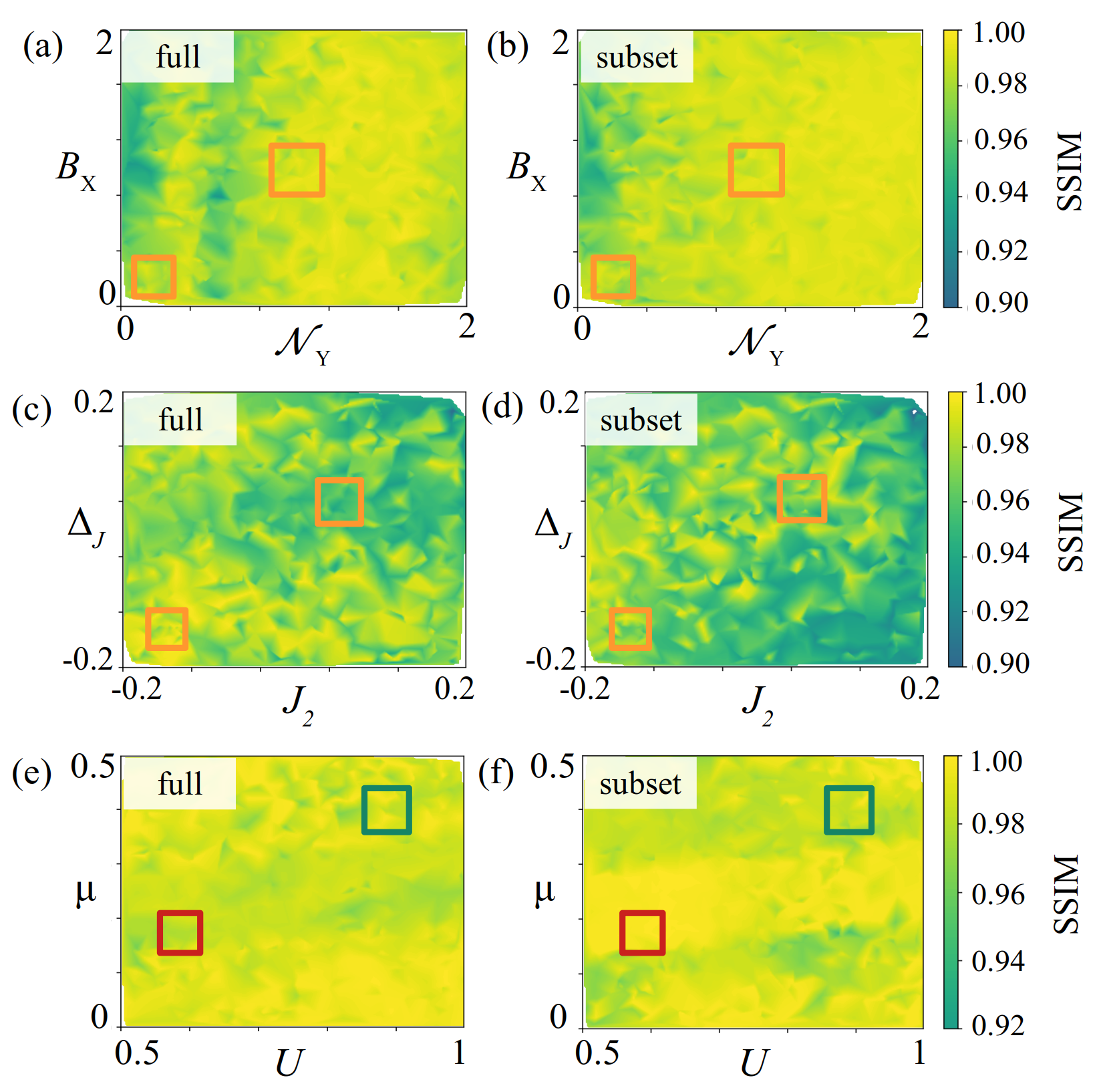}
    \caption{Similarity maps of the considered parameterspace for the 
    $S=1/2$ model (a,b),
    the     $S=1$ model (c,d)
    and the     Hubbard model (e,f).
    Panels (a,c,e) are obtained comparing the similarity
    between different real data and the cGAN trained on the full parameterspace. 
    Panels (b,d,f) are obtained comparing
    real data with the cGAN trained on a restricted subset.
    The training of the restricted cGAN was performed excluding
    examples inside the orange (and red) region.
    It is observed that the similarity between both cGANs does not show significant differences besides small fluctuations due to the intrinsic randomness.
    }
        \label{fig:ssim_map}
\end{figure}
We now partition the training and test set as depicted in Fig. \ref{fig:train_val} in App.~\ref{app:C} where the cases contained in the orange squares are excluded from the training. Afterward, the performance of the cGAN is computed in the whole phase space (Fig. \ref{fig:ssim_map}~(b,d,f)).
The measure of the similarity between two images is in this case the structural similarity index measure (SSIM)~\cite{Wang2004,Dosselmann2009,Wang,Brunet2012}. In contrast to other techniques e.g. the MSE, the SSIM does not compute absolute errors. It compares the structural information of two given images taking into account spatial correlations among pixels. The SSIM is defined in the interval between 0 and 1, where 1 is the maximum, only reached for two identical images.
As a reference, in Fig. \ref{fig:ssim_map}~(a,c,e) we are comparing real data with different hidden variables to generated data of the cGAN trained on the full parameter space. In Fig. \ref{fig:ssim_map}~(b,d,f), real data with random hidden variables is compared with generated data of the cGAN trained on a restricted subspace.
As shown in Fig. \ref{fig:ssim_map}~(b,d,f), the similarity index between real and generated dynamical correlators shows a high value over the whole parameter space, quantitatively similar to Fig.~\ref{fig:ssim_map}~(a,c,e). Most importantly it is observed that the cGAN trained in the restricted space (Fig. \ref{fig:ssim_map}~(b,d,f)) shows similar performance inside and outside the training region even in direct comparison with the cGAN trained on data including the restricted area (Fig. \ref{fig:ssim_map}~(a,c,e)). This highlights that the cGAN, restricted to a subset of the parameter space, is capable of generating data outside its trained region. Another reason for the different landscapes of the similarity maps is the cGAN training process where the weights are randomly initialized. This always leads to a slightly different training outcome which in this case can be seen in slight differences in the similarity maps. This, however, does not affect the averaged accuracy over the full parameterspace.

In App.~\ref{app:C}, we study the SSIM map in the vicinity of the validation areas in more detail (see Fig.~\ref{fig:ssim_map_zoom}). To summarize the results, we observe that the restricted cGAN gives results whose similarity is analogous to the cGAN trained on the full parameterspace. Most importantly, we observe that the SSIM takes similar values inside and outside the excluded areas, indicating that the cGAN learns to faithfully generate data in parts of the phase diagram not used for the training.

\begin{figure}[t!]
    \includegraphics[width=.99\columnwidth]{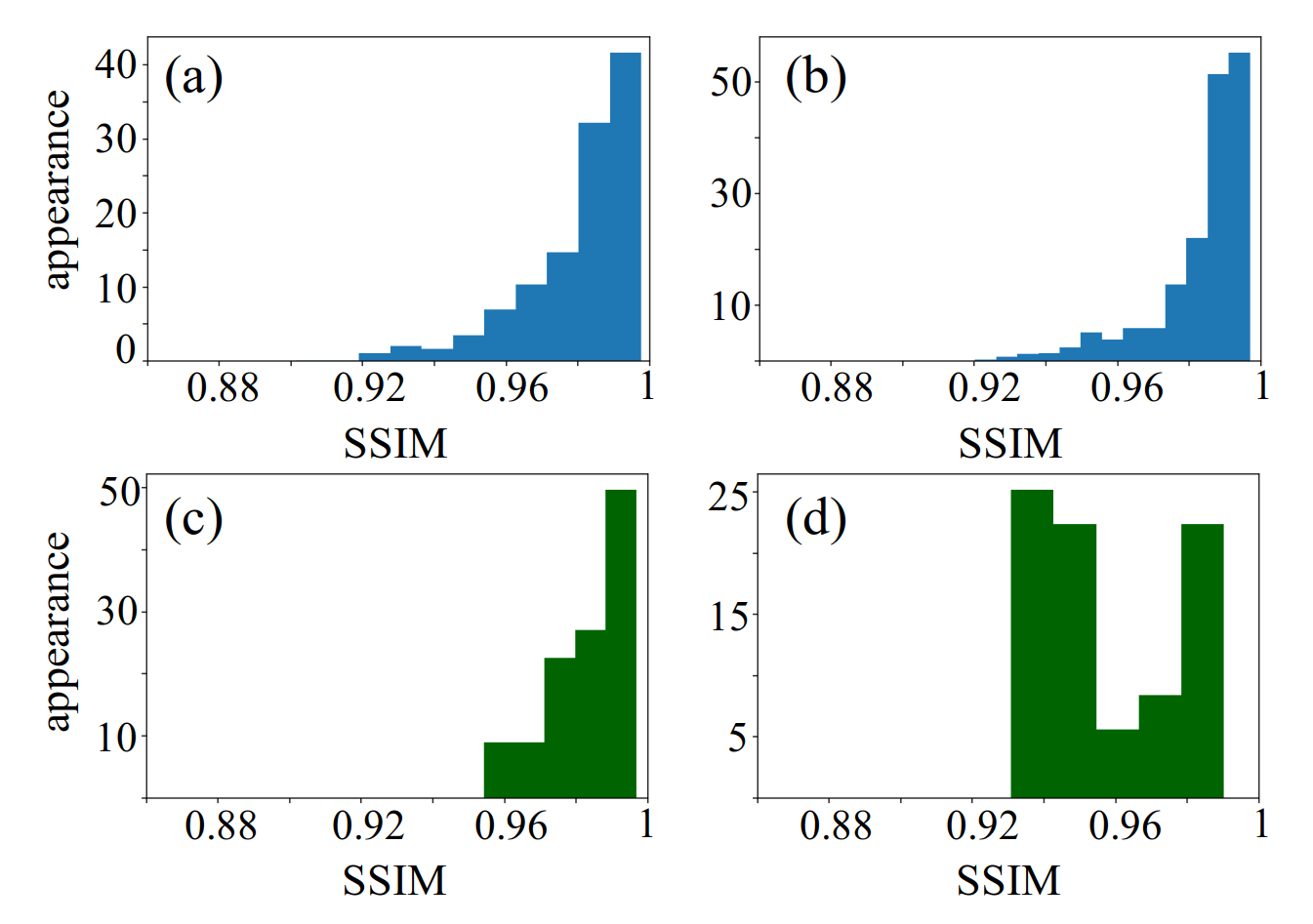}
    \caption{
    Statistics of the SSIM for the $S=1/2$ model.
    Panels [(a) and (c)] is the cGAN trained on the total parameter space,
    panels [(b) and (d)] correspond to cGAN trained is the restricted space. 
    Panels (a) and (b) are the SSIMs for the whole parameter space (randomly sampled for 400 parameter combinations) and panels (c) and (d) are the SSIMs for the validation set only.
    It is observed that the two cGANs have nearly the same performance,
    highlighting the extrapolation capability of the restricted cGAN.
    }
        \label{fig:hist_s12}
\end{figure}

\begin{figure}[t!]
    \includegraphics[width=.98\columnwidth]{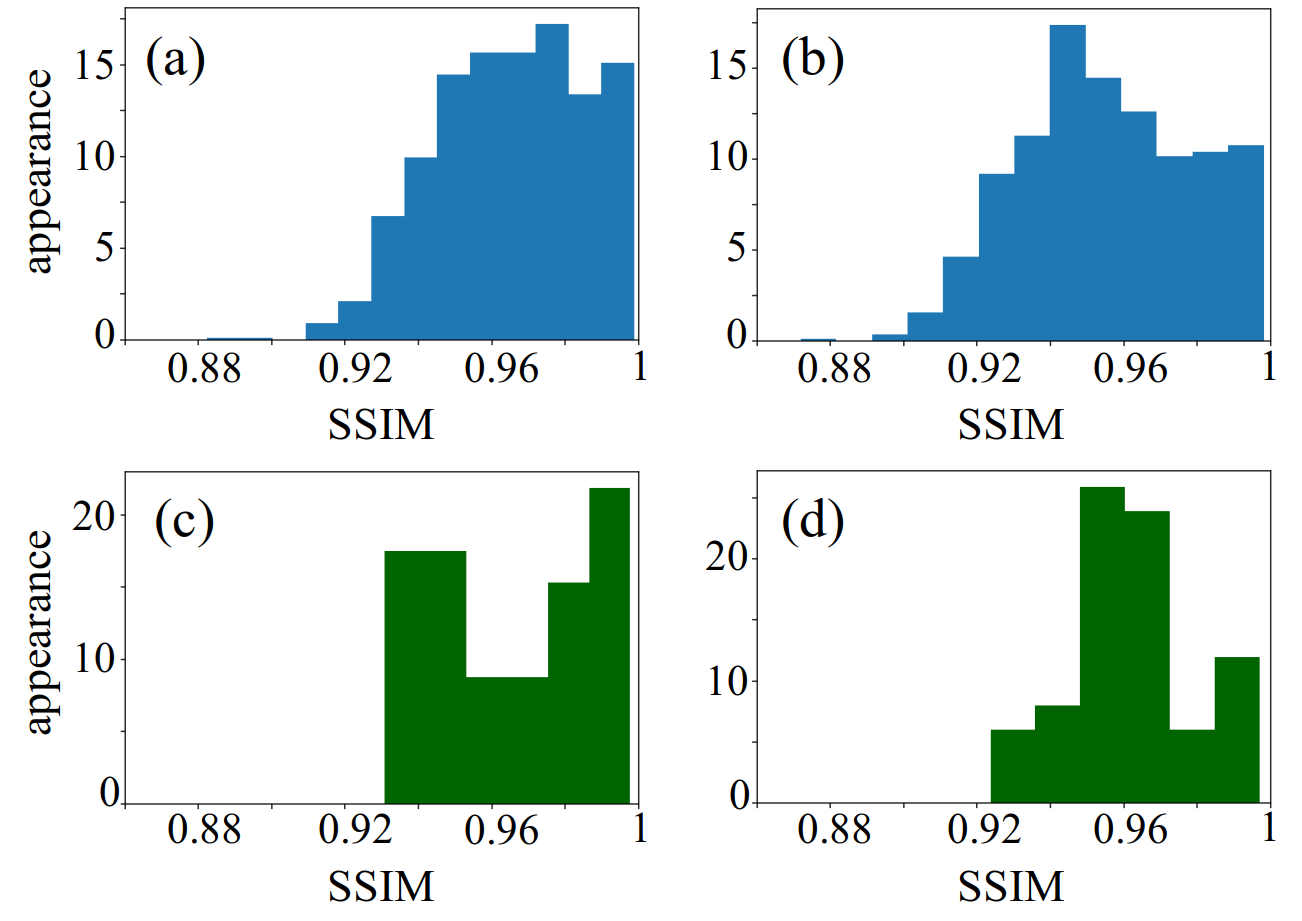}
    \caption{
    Statistics of the SSIM for the $S=1$ model.
    Panels [(a) and (c)] is the cGAN trained on the total parameter space,
    panels [(b) and (d)] correspond to cGAN trained is the restricted space. 
    Panels (a) and (b) are the SSIMs for the whole parameter space (randomly sampled for 400 parameter combinations) and panels (c) and (d) are the SSIMs for the validation set only.
    It is observed that the two cGANs have nearly the same performance,
    highlighting the extrapolation capability of the restricted cGAN.
    }
        \label{fig:hist_s1}
\end{figure}

\begin{figure}[t!]
    \includegraphics[width=.999\columnwidth]{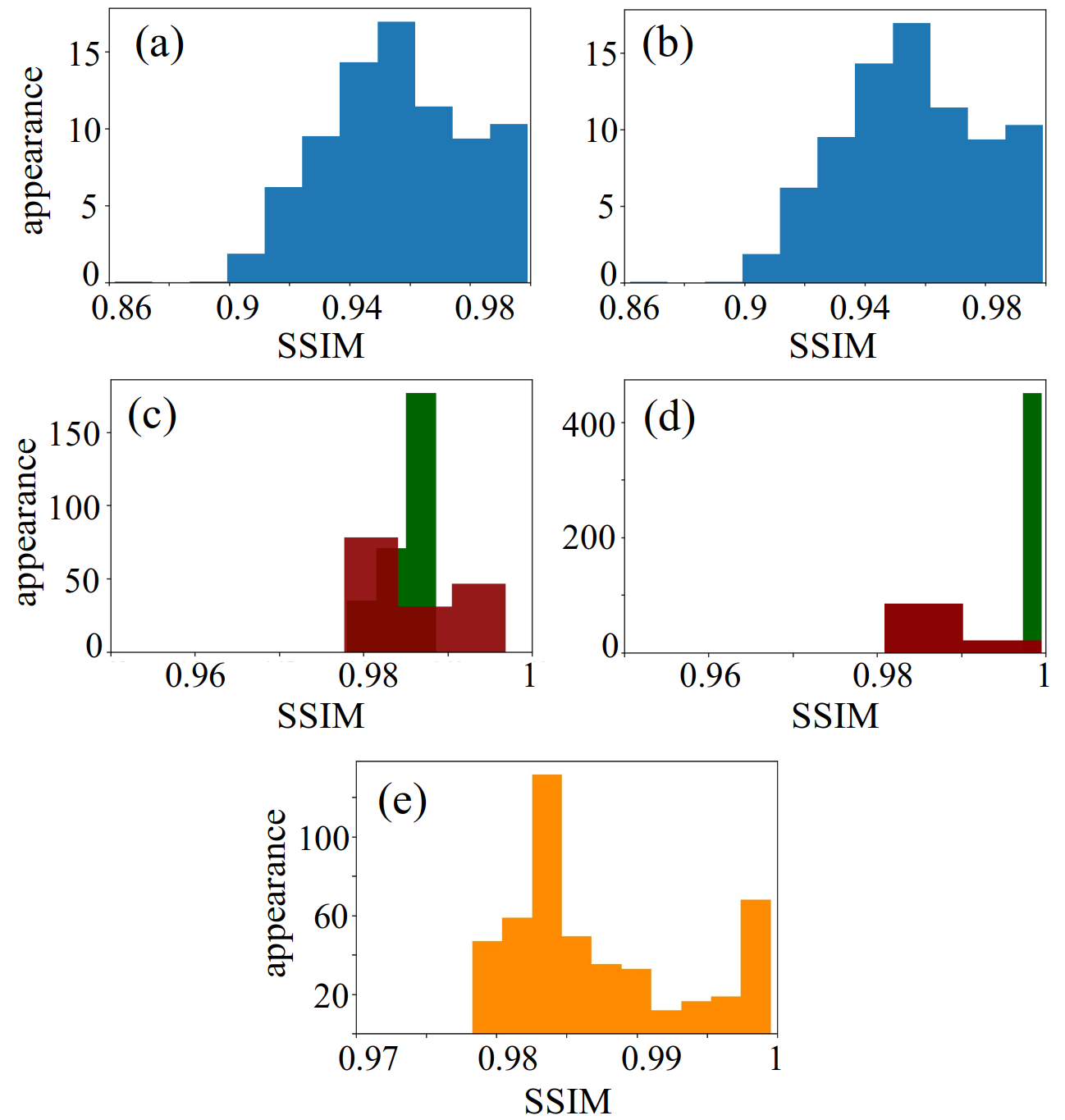}
    \caption{
    Statistics of the SSIM for the Hubbard model.
    Panels (a) and (c) is the cGAN trained on the total parameter space,
    panels (b) and (d) correspond to cGAN trained is the restricted space. 
    Panels (a) and (b) are the SSIMs for the whole parameter space (randomly sampled for 400 parameter combinations) and panels (c) and (d) are the SSIMs for the validation set only.
    The two colors in (c,d) correspond to the two different areas excluded.
    It is observed that the two cGANs have nearly the same performance,
    highlighting the extrapolation capability of the restricted cGAN.
    Panel (e) shows the SSIM for a single parameter, averaged over different
    200 different generated images of the cGAN,
    showing that the similarity changes with each simulation of the cGAN due to the 
    intrinsic randomness.
    }
        \label{fig:hist_H}
\end{figure}

We now focus on discussing the statistical analysis of the benchmarks. We show in Fig.~\ref{fig:hist_s12} the analysis for the $S=1/2$ chain, in Fig.~\ref{fig:hist_s1} the analysis for the $S=1$ chain, and in Fig.~\ref{fig:hist_H} the analysis for the Hubbard model.
In particular, we compare the similarity index obtained for the network trained in the full space (Fig.~\ref{fig:hist_s12}~(a,c), Fig.~\ref{fig:hist_s1}~(a,c), and Fig.~\ref{fig:hist_H}~(a,c)), and in the restricted space (Fig.~\ref{fig:hist_s12}~(b,d), Fig.~\ref{fig:hist_s1}~(b,d), and Fig.~\ref{fig:hist_H}~(b,d)). It is clearly observed that, for the three considered models, the two cGANs have nearly the same performance, highlighting the extrapolation capability of the restricted cGAN. This finding clearly shows that cGAN effectively learns to extrapolate the dynamical data. 
This is clearly seen in Fig.~\ref{fig:val_pred} in App.~\ref{app:C}, where we show a comparison between a real dynamical correlator and a generated one, in a point in the phase space in which the network was not trained, showing a faithful agreement.
Finally, to emphasize the role of hidden parameters, we show in Fig. \ref{fig:hist_H}(e) the similarity index between generated images with different randomness. This further shows that the small differences between images with the same parameters are well captured by the cGAN, and in particular have the same order of magnitude of fluctuations as the SSIM in the whole phase space.

\section{Hamiltonian inference and data assessment with the generative model}
\label{sec:disc}

\begin{figure}[t!]
    \includegraphics[width=.99\columnwidth]{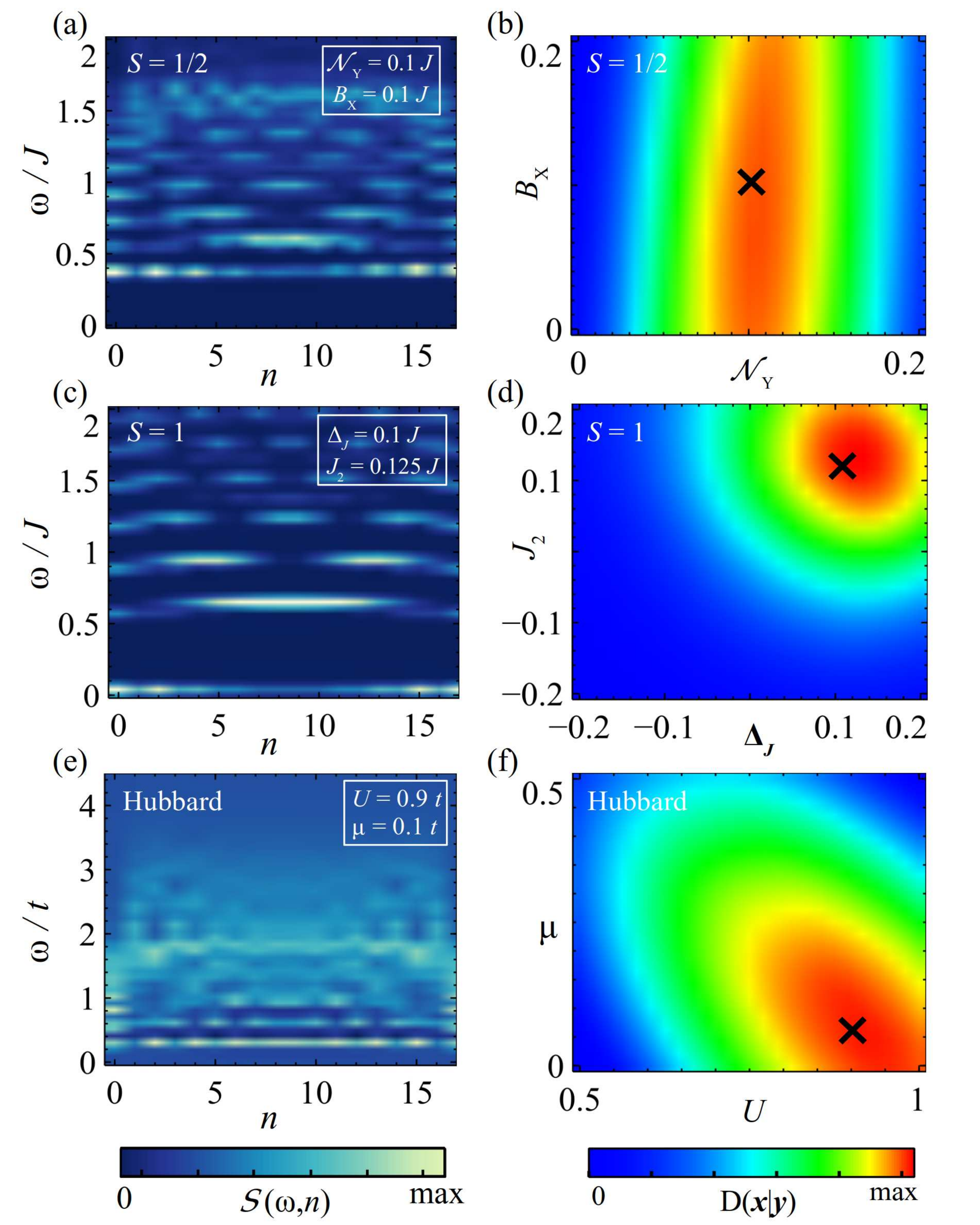}
    \caption{
    Hamiltonian learning with the discriminator network of the cGAN for the many-body spectra of a one-dimensional $S=1/2$ system (a), $S=1$-system (c), and the Hubbard model (e). The dynamical correlators $S(\omega,n)$ are defined in Sec.~\ref{sec:many}. The corresponding predictions of the probabilities of the discriminator, defined as D$(\textbf{x}|\textbf{y})$ (see Eq.\ref{Eq:val2}) with $\textbf{x}$ as input spectra and $\textbf{y}$ as conditional parameters, are shown in (b,d,f) in a squared scale. The black crosses mark the exact location of the conditional parameters of the input spectra.
    }
        \label{fig:disc_d3}
\end{figure}

In this section, we demonstrate how our conditional generative adversarial model allows us to tackle two additional tasks by exploiting the trained discriminator network as an automatic byproduct of the trained algorithm. The focus of this section is not the generator that was responsible for the generation of the spectra in the previous sections, but the discriminator which is to this point only used during the cGAN training process. The discriminator is trained to distinguish between real, physical systems and unrealistic ones. This feature provides the fundamental ingredient to perform parameter inference and anomaly detection.

\subsection{Hamiltonian learning with the generative model}
Here we show how the discriminator network of the generative model allows to directly extract the physical parameters of a certain dynamical correlator. The estimation of physical parameters from data is commonly referred to as Hamiltonian learning and has been explored with a variety of machine learning techniques~\cite{PhysRevLett.112.190501,PhysRevA.89.042314,Wang2017,PhysRevLett.124.160502,PhysRevResearch.1.033092,2021arXiv210301240V}. While these methodologies are usually specifically developed for this purpose, conditional generative algorithms provide this functionality as a direct consequence of their training.

\begin{figure*}[t!]
    \includegraphics[width=1.35\columnwidth]{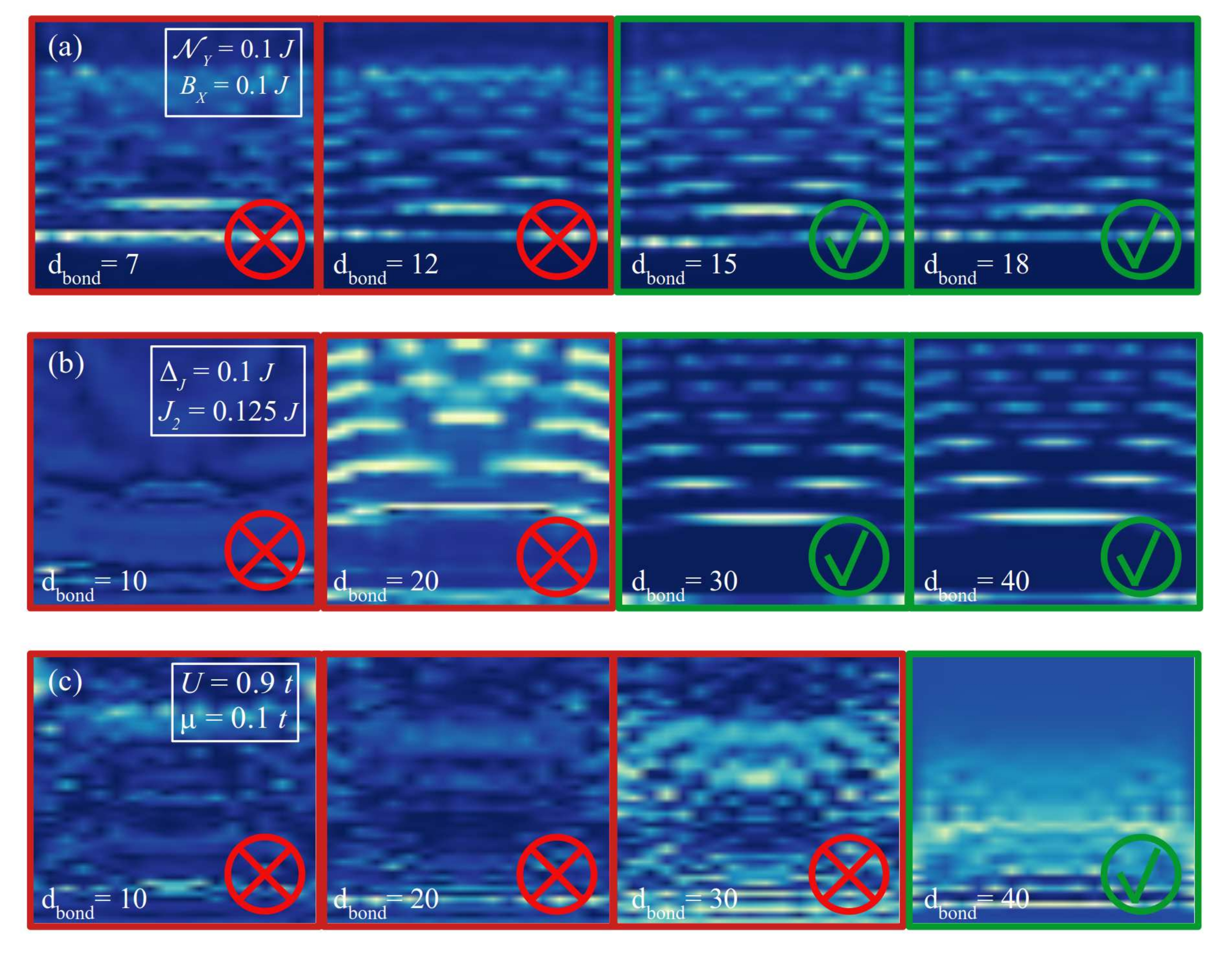}
    \caption{Outlier detection of faulty unphysical many-body dynamical spectra, emulated by using insufficient bond dimension in tensor-network calculations for the $S=1/2$ (a), $S=1$ (b), and Hubbard model (c). The green rectangle marks the bond dimension for which the dynamical correlator is considered as real by the discriminator of the cGAN.}
        \label{fig:disc_outlier}
\end{figure*}

The discriminator learns to assess if a certain dynamical correlator corresponds to the physical parameters given as conditional inputs. Due to instantaneous efficiency, this functionality can be directly applied to inquire the discriminator if a dynamical correlator corresponds to every single possible Hamiltonian. This procedure allows extracting the confidence that the discriminator has for the whole set of parameters as shown in Fig.~\ref{fig:disc_d3}. The probability estimations are shown in Fig.~\ref{fig:disc_d3}~(b), Fig.~\ref{fig:disc_d3}~(d), and Fig.~\ref{fig:disc_d3}~(f) for the three studied many-body systems. 

For the $S=1/2$ many-body dynamical correlator in Fig.~\ref{fig:disc_d3}~(a), the discriminator is able to predict the conditional parameter $\mathcal{N}_y$ with high accuracy, yet yielding a whole range for $B_x$. This is due to the fact that $\mathcal{N}_y$ has a very strong impact on the dynamical correlator where on the other hand $B_x$ leaves the spectrum almost the same. Therefore, the discriminator detects a strong dependence on $\mathcal{N}_y$ which appears in the exact tensor-network calculations. Interestingly, the consistency of that dynamical correlator with several Hamiltonians simultaneously would represent a challenge for parameter extraction purely based on supervised learning due to the non-unique parent Hamiltonian~\cite{Cao2020}, representing an advantage of generative-based parameter estimation.

We now move on to the gapped $S=1$ chain. The parameter predictions for the $S=1$ spectrum, shown in Fig.~\ref{fig:disc_d3}~(d), provide a single maximum in the parameter assessment,  determining the real parameter with good precision. In comparison with the $S=1/2$, for the $S=1$ the parameters are uniquely determined by the provided dynamical spin correlator. This enhanced accuracy can also be rationalized from the existence of both bulk and edge excitations, that provide potentially complementary information to the discriminator.

Finally, we move on to the interacting fermionic Hubbard model. The predictions of the Hubbard model, shown in Fig.~\ref{fig:disc_d3}~(f), estimate the area of the exact conditional parameters well and provide a unique maximum for the estimated parameter. In comparison with the $S=1$ model, the estimated area is larger and less steep, which can be related to the higher complexity of the spectra of the many-body model. In particular, the features of the spectra are not as clear and distinct as it is the case of the $S=1$ model. Considering that we use the same amount of training data for each many-body system, the differences in accuracy may be related to the higher complexity of the dynamical correlator of the Hubbard model in comparison with the $S=1$ chain. 

Here, we elaborate on the advantage of the cGAN over traditional methods to estimate the phase space of a Hamiltonian. From the theoretical point of view, computing the dynamical correlator requires solving an interacting quantum many-body Hamiltonian. As a reference, computing the dynamical correlator of a $S=1/2$ chain with exact dense diagonalization takes 1 second for 10 sites, 10 hours for 15 sites, and it would take 900 years for 20 sites. Many-body tensor-networks~\cite{PhysRevLett.69.2863,RevModPhys.77.259,Verstraete2008,Schollwck2011} provide a dramatic speed up with respect to exact diagonalization, allowing to compute the dynamical correlator of an $S=1/2$ with 20 sites in 30 minutes. However, when considering more complex models such as the Hubbard
model, computing the dynamical correlator~\cite{PhysRevB.60.335,PhysRevB.90.045144,Schollwck2011} for 20 sites increases already to 10 hours. All those times above correspond to a single Hamiltonian. In practice, if we are interested in exploring the full phase diagram of a Hamiltonian, computing dynamical correlators even with tensor networks quickly becomes unfeasible.
This becomes especially critical for procedures that unavoidably involve exploring the full combination of parameters in the Hamiltonians, as is the case of Hamiltonian learning. 
In particular, with the cGAN algorithm demonstrated in our manuscript, we can generate dynamical correlators of a given Hamiltonian almost instantaneous. Under consideration of the required training data, the cGAN approach offers an significant speedup over the previous mentioned computational methods, especially for studies of a huge region of the parameter space.
This point is explained in detail below. In practice, this implies that the Hamiltonian inference problem that requires 10 days with many-body tensor-networks, can be performed with our cGAN algorithm in 30 seconds.
This dramatic speed up demonstrates that cGAN are high valuable algorithms
to study quantum many-body systems, for a task that can be directly applied to experimentally measured data.

We would like to note that the existence of a wide confidence region is a direct consequence of the nature of our problem. In particular, the  models we are considering have a set of known parameters, acting as conditional parameters of the GAN, and most importantly a set of unknown parameters to the algorithm that enter as the generative noise of the GAN.
In particular, this is directly observed in Fig.~\ref{fig:ssim_map}(a,c,e), that correspond to the similarity between real images with different hidden variables.
Physically, this implies that we are focusing on systems in which we know some physical parameters, but that potentially contain a variety of hidden variables not considered experimentally and that can potentially slightly modify the final result. This would be the typical situation in an actual experiment, in which the hidden parameters account for disorder, extra terms in the Hamiltonian, or perturbations to the systems that are not directly controllable.
In other words, experimentally Hamiltonians can have extra terms not originally considered in the model, and this uncertainty is explicitly incorporated in our algorithm through the noise of the generative network. As a result, the confidence in the parameter extraction of Fig.~\ref{fig:disc_d3} directly reflects the fact that the hidden variables can slightly impact the dynamical correlators of the system, giving rise to an intrinsic uncertainty when doing the parameter extraction. In particular, in the absence of the generative noise, the confidence of Fig.~\ref{fig:disc_d3} would be much narrower, as no hidden variables would be included in the model.

For the three considered, models the accuracy of the estimation shows small variations across the different parameter realization and noise level, yet overall giving a good estimation of the vicinity of the exact conditional parameters for each of the studied systems. While we performed here parameter extraction solely with the dynamical correlators, it is worth noting that an analogous procedure can be extended by training a generative model with combined time-dependent~\cite{PhysRevLett.127.150504,2021arXiv210301240V} or ground state observables~\cite{PhysRevX.8.021026,PhysRevX.8.031029}. Finally, it is worth noting that while here we focused on simulated dynamical correlators, this procedure can be readily applied with experimentally measured spin excitations~\cite{Spinelli2014,Toskovic2016}, providing a procedure for experimental Hamiltonian extraction with conditional generative adversarial networks.    

\subsection{Generative model as a many-body assessor}
When observing complex phenomena in a quantum system, a key question is if the observed behavior corresponds to the targeted physical state of the system or reflects an undesired artifact of the underlying methodology or setup~\cite{2021arXiv211010378C,Baireuther2018,Baireuther2019}. In particular, many-body calculations often require a degree of controlled accuracy that is model and method-dependent. However, estimating if a certain many-body phenomenon represents a physical system solely from the observation of the dynamical excitations represents an outstanding challenge even for human experts. Here we address how the discriminator provides a direct algorithm to assess if a certain dynamical correlator corresponds to a physically-meaningful system, or rather reflects an artifact in the underlying methodology.

The trained discriminator can directly assess if a certain input corresponds to a real result, as a direct consequence of the competitive training with the generator. This procedure provides a discriminator-based outlier detection at no cost after the training of the generative model.
To study this effect, we generate different dynamical correlators computed with different degrees of accuracy, which in our tensor-network formalism is directly controlled by the bond dimension of the matrix product state. The discriminator is then used to detect spectra with insufficient accuracy corresponding to ill-converged results. Figure~\ref{fig:disc_outlier} shows the results of the outlier detection for each many-body system. For the $S=1/2$ system in Fig.~\ref{fig:disc_outlier}~(a), the discriminator detects outliers for a bond-dimension lower than $d_{\text{bond}}=15$, for the $S=1$ system the numerical accuracy becomes insufficient below $d_{\text{bond}}=30$, and for the Hubbard model, the bond dimension has to be close to $d_{\text{bond}}=40$, in order to be considered a physically meaningful result by the discriminator. The increasing bond dimension required to pass the discriminator test is a direct consequence of the increasing local Hilbert space of the underlying models and reflects the higher entanglement of the respective many-body states.

Here, we elaborate on the reason behind the degradation of the data as $d_{bond}$ is lowered. The tensor-network calculations were performed with bond dimension $d_{bond}$, which is high enough to provide fully-converged dynamical correlators for three models considered. As the bond dimension is lowered, the quality of the calculation gets worse, yet the worsening is model-dependent. In particular, for the $S=1/2$ model (Fig.~\ref{fig:disc_outlier}~(a)) a bond dimension $d_{bond}=30$ still provides dynamical correlators with reasonable quality, whereas for the Hubbard chain (Fig.~\ref{fig:disc_outlier}~(c)) the quality decreases much faster as the bond dimension is decreased. 
A more detailed study of the Hubbard model (Fig.~\ref{fig:disc_outlier}~(c)) in the bond-dimension range $d_{\text{bond}}=30-40$ is provided in App.~\ref{app:C}.

It is worth noting that all previous assessment is performed including noisy $\xi^\alpha$ terms in the Hamiltonian, demonstrating that the generative model distinguishes between physical noise in the Hamiltonian parameters and artifacts stemming from the computational procedure. Furthermore, while in this section we focused on simulated data, an analogous procedure can be extended to experimental data, providing a methodology to assess experimental measurements using generative adversarial learning.

For the models we considered in this work, we have not explored the possibility of using the GAN algorithm to extract the critical points of the model. This would certainly be a possibility of great interest in the future. From the  practical point of view, a procedure analogous to Nature Physics 13, 435 (2017) could be implemented with the dynamical correlators using learning by confusion. Let us consider the specific case of a phase transition for a quantum-disordered magnet to a magnetically ordered state. In this situation, the magnet would host magnon excitations~\cite{Giamarchi2003} that have a well-defined energy versus momentum relation, which would directly be reflected in the dynamical correlator. In stark contrast, for a quantum disordered magnet, spin excitations would correspond two-spinon modes~\cite{Giamarchi2003} that lead to a continuum of $S=1$ excitations that lack a well-defined energy-momentum dispersion. This qualitative difference between the two phases directly signals the nature of the ground state. We note that this is analogous to the magnetization observable used in Nature Physics 13, 435 (2017). As a result, using the strategy of learning by confusion with the discriminator would automatically allow extracting the critical point.

\section{Conclusions}
\label{sec:summary}
To summarize, we demonstrated how continuous conditional generative artificial networks allow simulating dynamical correlators for many-body systems which are almost indistinguishable from exact many-body calculations. In stark contrast with conventional supervised algorithms, our methodology allows to simultaneously account for hidden variables unknown to the model, intrinsically account for randomness in the models, reduce by about one order of magnitude the required many-body data required for the training, and exploit the discriminator for Hamiltonian inference and anomaly detection. In particular, we have demonstrated our methodology with three different types of many-body Hamiltonians, starting with a gapless $S=1/2$ model featuring spinon excitations, an interacting system with topological order and topological boundary modes, and an interacting fermionic system at arbitrary electron fillings. After the training process, the cGAN algorithm is able to simulate these systems instantaneous for arbitrary combinations of conditional Hamiltonian parameters. 
Furthermore, the trained cGAN is not only able to simulate these systems with the generator, the trained discriminator can also be utilized for estimating the parameters of a Hamiltonian from data and for the detection of outliers and wrong-labeled data without the requirement of additional training. These two features can be directly extended with the trained algorithm to data in order to determine unknown underlying Hamiltonians, detect artifacts, and ultimately they can be directly applied to experimental data.
Furthermore, we achieved a significant increase in performance of Hamiltonian learning over conventional methods which demonstrates the power of cGANs to study quantum many-body problems very efficient.

Our results establish a first step towards exploiting generative adversarial machine learning to simulate and design many-body matter. Beyond the results demonstrated here, it is worth noting that the trained cGAN algorithm can be used as a tool in experiments in order to investigate underlying unknown Hamiltonians of systems, either with the generator or discriminator, considering the speed-up for computations of big system sizes and the possibility to simulate Hamiltonians with arbitrary parameters. It is also worth noting that our results use a fully-connected deep neural network for each the generator and discriminator, leaving plenty of room for further future optimizations with deep convolutional neural networks for image compression and feature extraction. Those improvements will allow increasing the accuracy of the cGAN, combining different systems into one single algorithm, and increasing the system size with pre-training on smaller systems which are computationally more feasible to generate.

\textbf{Acknowledgements}:
We acknowledge the computational resources provided by
the Aalto Science-IT project,
and the financial support from the
Academy of Finland Projects No. 331342 and No. 336243
and the Jane and Aatos Erkko Foundation.
We thank O. Zilberberg, T. Neupert, M. H. Fischer,
M. M. Denner, and E. Greplova for useful discussions.

\section*{Appendix}

\appendix

\section{GAN architecture} \label{app:A}

For each many-body system of Sec.~\ref{sec:GAN} we are training a separate cGAN. The explicit network architecture and training parameters can be found in Tables~\ref{tab:generator} and~\ref{tab:disc} as well as in the code~\cite{codecgan}. In general, we are using fully-connected deep neural networks for the generator and discriminator with a maximal layer dimension of 4048. For the hidden layers, we use the \textit{LeakyRelu} activation function~\cite{Maas2013RectifierNI}. Kernel regulizers are included in every dense layer applying penalties on layer parameters during the training. The output activation function of the discriminator is the \textit{sigmoid} function, and of the generator the output function is the tanh-function~\cite{goodfellow2014generative}. Additional Gaussian noise added into the discriminator (see Table~\ref{tab:disc}) helps to stabilize the training process.\vspace{0.5mm} 

\textit{Training -} We are training a separate cGAN for each system. We are using a batch size of 100 and 40 epochs for the training. Choosing a smaller batch size of 50 for 5-10 additional epochs may increase the accuracy of the networks. For the generator, we are using the Adam optimizer with a learning rate of 0.001 and for the discriminator stochastic gradient decent. The images of the real space excitation spectra are flattened into a 1D array of size 900 as input for the fully-connected neural network of the generator.

\begin{table}[htbp] 
\centering
\renewcommand{\arraystretch}{0.5}
    \begin{tabular}{| c | c | c | }\hline
        \hspace{0.5cm} Layer \hspace{0.5cm} & \hspace{0.5cm} Activation \hspace{0.5cm} & \hspace{0.5cm} Dimension \hspace{0.5cm} \\[0.5ex]\hline \hline
        input  & concatenation & $[10,2] \rightarrow 12$\\[0.5mm]
        dense  & LeakyReLu & 2024 \\[0.5mm]
        dense  & LeakyReLu & 4048 \\[0.5mm] 
        dense  & Tanh & 900 \\[0.5mm]\hline
    \end{tabular}
    \caption{Architecture of the generator network.}
    \label{tab:generator}
\end{table}

\begin{table}[htbp] 
\centering
\renewcommand{\arraystretch}{0.5}
    \begin{tabular}{| c | c | c | }\hline
        \hspace{0.5cm} Layer \hspace{0.5cm} & \hspace{0.5cm} Activation \hspace{0.5cm} & \hspace{0.5cm} Dimension \hspace{0.5cm} \\[0.5ex]\hline \hline
        input  & concatenation & $[900,2] \rightarrow 902$\\[0.5mm]
        input  & Gaussian noise & 902 \\[0.5mm]
        dense  & LeakyReLu & 4048 \\[0.5mm]
        dense  & LeakyReLu & 2024 \\[0.5mm] 
        dense  & sigmoid & 1 \\[0.5mm]\hline
    \end{tabular}
    \caption{Architecture of the discriminator network.}
    \label{tab:disc}
\end{table}

\section{Training data generation} \label{app:B}
For each many-body system, we created 2250 spectra with random conditional parameter combinations. We calculated the dynamical correlators as described in Sec.~\ref{sec:dynamics} for the many-body Hamiltonians introduced in Sec.~\ref{sec:many}. In order to minimize the required calculations for the creation of the training set, we are using two methods to enhance the number of examples without doing further tensor-network calculations. First, we are mirroring the 1d systems of 18 sites around site 9 and therefore double the number of systems. Second, we are mimicking the background noise of the random parameters without affecting the conditional parameters. We are changing the intensities depending on the frequency regime by defining a function that adds fluctuations around each energy level seen in the spectra of Sec.~\ref{sec:many}. The data augmentation function is defined as 
  \begin{equation}
      f(\omega) = \omega \left[ 1 + \xi_1 \text{cos}(\xi_2 \omega) \right]
  \end{equation}
with random numbers $\xi_1 \in [0.04, 0.1]$ and $\xi_2 = 2\pi * n / 2.5$ with  $n \in [1, 4]$. This function adds a different amount of intensity to the dynamical correlator depending on the frequency $\omega$. In an experiment, this procedure would mimic any potential form factor that would affect the intensity of a many-body transition intrinsic to the measurement setup. This method allows adding an arbitrary amount of training examples with the same conditional parameters but with different noise values. In our case, we could enlarge the number of systems from 2250 original tensor-network calculations to 36 000 which represent our full training set. The pre-processing procedure can be found in~\cite{codecgan} and consists of a re-scaling of the input spectra between 0 and 1, as well as a separate re-scaling of the conditional parameters.

\section{Further analysis of the cGAN performance} \label{app:C}

\subsection{Generator}

\begin{figure}[h]
    \includegraphics[width=.99\columnwidth]{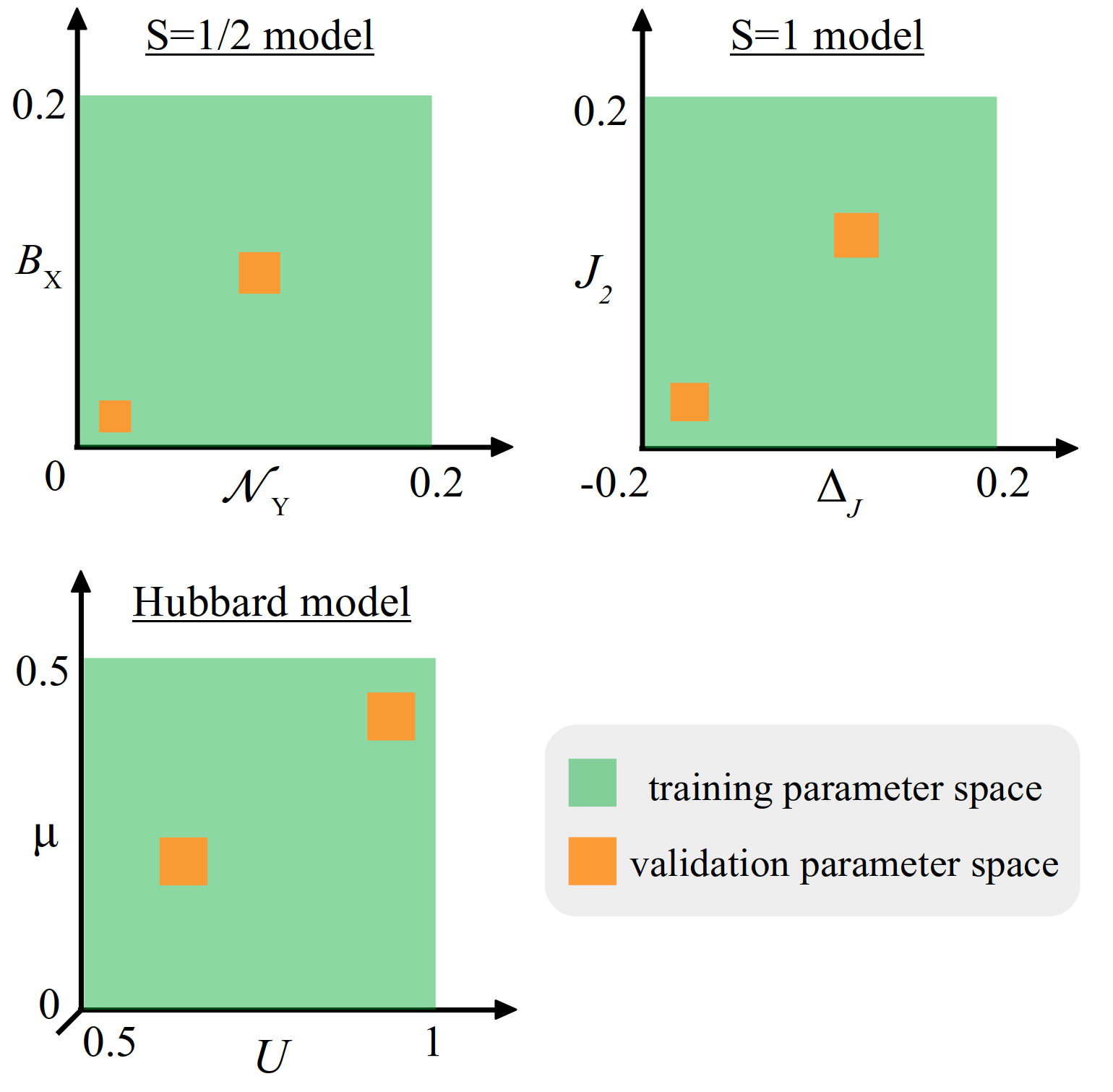}
    \caption{Training-validation split in the full parameter range.
    The orange areas correspond to regions for which no example is
    provided to the cGAN algorithm. The orange regions were chosen
    randomly and no strong dependence on the results is found with respect
    to their location.
    }
        \label{fig:train_val}
\end{figure}

\begin{figure}[h]
    \includegraphics[width=.97\columnwidth]{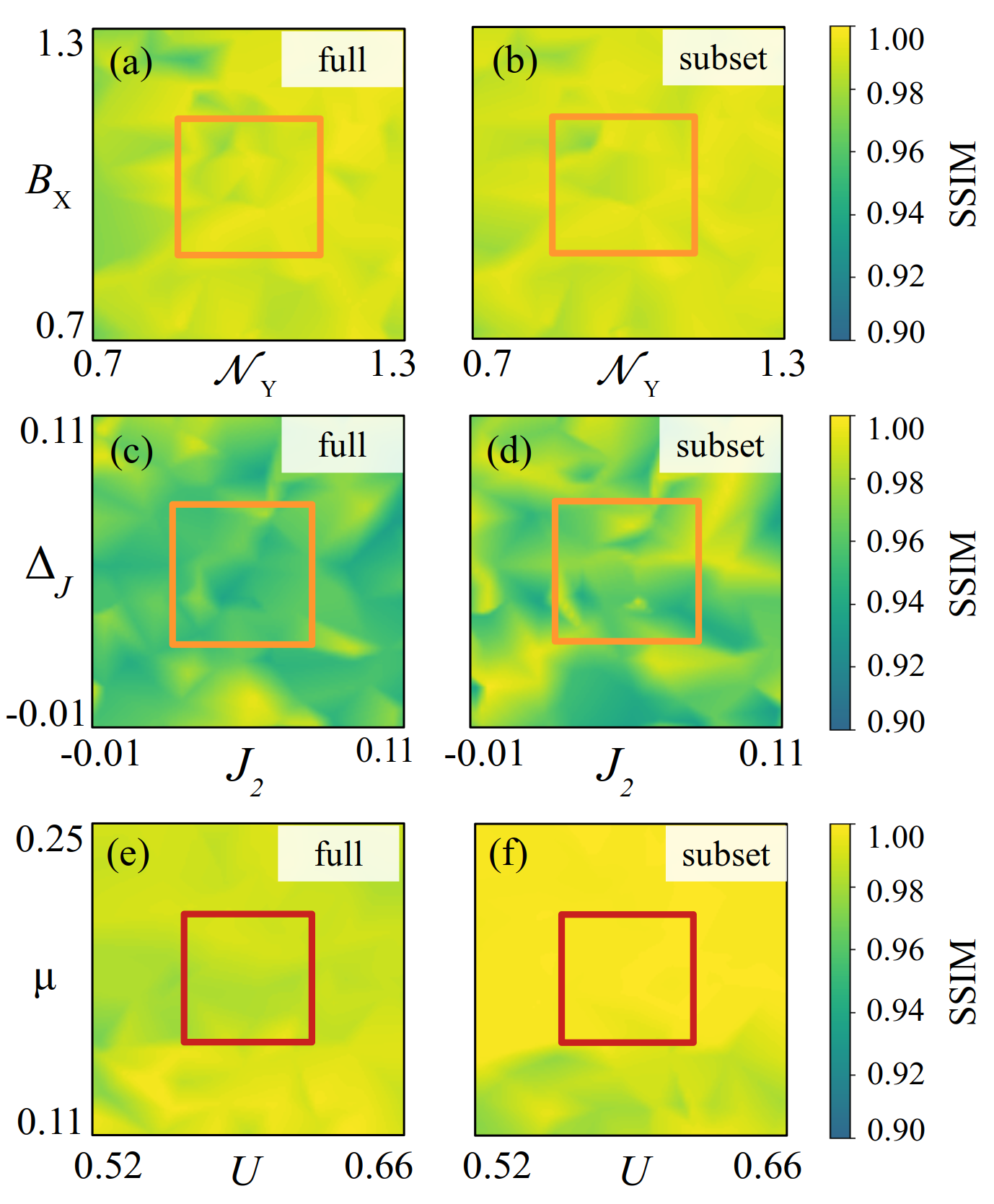}
    \caption{Similarity maps for the  $S=1/2$ model (a,b), the $S=1$ model (c,d) and the Hubbard model (e,f). Panels (a,c,e) are obtained comparing the similarity between different real data and the cGAN trained on the full parameter space. Panels (b,d,f) are obtained comparing real data with the cGAN trained on a restricted subset.
    The training of the restricted cGAN was performed excluding examples inside the orange (and red) region.
    It is observed that the similarity between both cGANs does not show significant differences besides small fluctuations due to the intrinsic randomness.
    The quality of the generations of both cGANs are furthermore analogous
    inside and outside the region excluded in the training.
    }
        \label{fig:ssim_map_zoom}
\end{figure}

As a first demonstration, we show in Fig.~\ref{fig:ssim_map_zoom}~(a) quantitative benchmark of the cGANs, one trained in the full parameter space and one trained in a restricted subset. 
The quality factor compares the similarity between real data with random hidden variables and the cGAN generated data with the cGAN trained on the full parameter space (Fig.~\ref{fig:ssim_map_zoom}~(a,c,e)), and trained on a subset of parameters (Fig.~\ref{fig:ssim_map_zoom}~(b,d,f)).
 The parameter space inside the orange and red squares is not included in the training for the restricted cGAN (see also Fig.~\ref{fig:train_val}), but it is included in the validation to benchmark the extrapolation capabilities of the network.
In particular, we compute

\begin{equation}
    D^{\text{full}} = 
    \mathcal{D} (\mathcal{S}^{\text{real}} (\lambda,\chi)- \mathcal{S}^{\text{cGAN}}_{full}(\lambda))
\end{equation}

and 

\begin{equation}
    D^{\text{subset}} = 
    \mathcal{D} (\mathcal{S}^{\text{real}} (\lambda,\chi)- \mathcal{S}^{\text{cGAN}}_{subset}(\lambda))
\end{equation}

where $\mathcal{S}^{\text{real}}$, $\mathcal{S}^{\text{cGAN}}$ are the dynamical correlators computed with the tensor-network formalism and the cGAN, $\chi$ is a set of random hidden variables and $\lambda$ are the physical parameters. The functional $\mathcal{D}$ measures the similarity between two images, and is taken as the structural similarity index measure (SSIM)~\cite{Wang2004,Dosselmann2009,Wang,Brunet2012}. The similarity between two sets of data is therefore given by the SSIM, where the closer the value to 1 the greater the agreement. We elaborate on the SSIM below.

The maps of Fig. \ref{fig:ssim_map_zoom}~(a,c,e) show the $D^{\text{full}}$ for the three models of this manuscript, and the maps of Fig.~\ref{fig:ssim_map_zoom}~(b,d,f) show $D^{\text{subset}}$ for each model.
In particular, the cGAN trained on the full parameter space in Fig.~\ref{fig:ssim_map_zoom}~(a,c,e) shows similar small fluctuations between both images compared to the subset of Fig.~\ref{fig:ssim_map_zoom}~(b,d,f). The fluctuations appear due to the hidden variables and should therefore be taken as the similarity of reference to benchmark the subset cGAN in the unknown parameter space inside the squares. For each point of the phase space, the values of the hidden variables are taken randomly for the computed Hamiltonian.
The comparison of the similarity measure inside the unknown parameter space highlights the extrapolation capability of the cGAN. It is observed that the restricted cGAN gives results whose similarity is analogous to the full-trained cGAN. Most importantly, it is observed that the SSIM takes similar values inside and outside the excluded areas, signaling that the cGAN learns to faithfully generate data in parts of the phase diagram not used for the training. 

\begin{figure}[t]
    \includegraphics[width=.99\columnwidth]{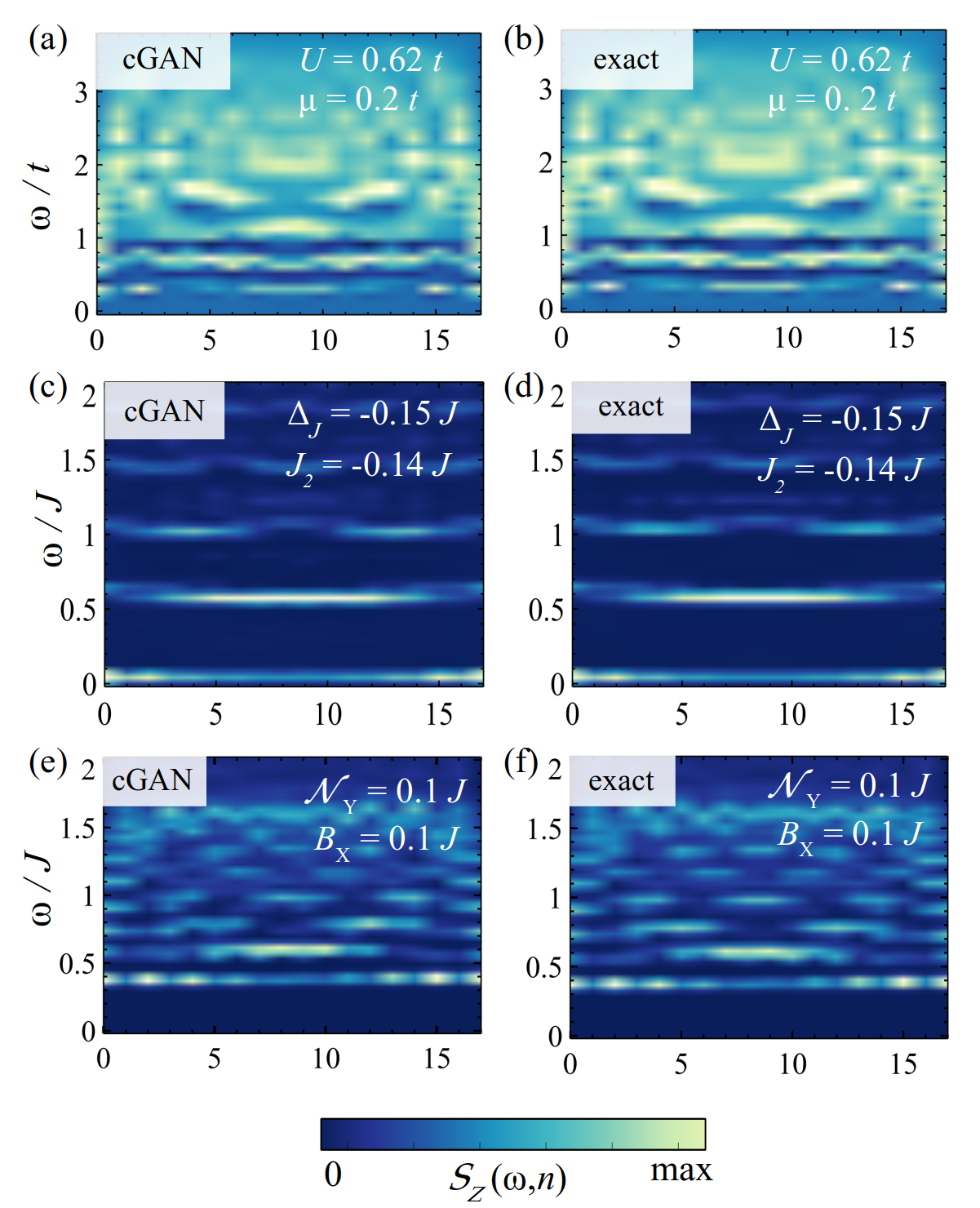}
    \caption{Examples for predictions on the validation set, for the
    network trained in the subspace. it is clearly observed that the
    generated images faithfully correspond to the real ones even
    for the phase space in which the network was not trained.
    }
        \label{fig:val_pred}
\end{figure}

We now elaborate on the reason for choosing the structural similarity index measure instead of a plain pixel-wise difference. Given the existence of diversity in the dynamical spectra for a set of parameters due to the hidden variables, spectra corresponding to the same parameters will have small differences. These small differences include small shifts in the peaks, and are a direct consequence of the hidden variables. A pixel-wise difference would yield a huge error even for images whose difference between peaks is very small, and therefore would not be a faithful measure of the similarity between these images.
An algorithm accounting for the similarity between images must therefore not consider such small shifts as a source of error, and rather focus on the overall features of the image. The structural similarity index~\cite{Wang2004,Dosselmann2009,Wang,Brunet2012} allows measuring the similarity between two images incorporating perceptual phenomena~\cite{Brunet2012}, and in particular focusing on the spatial correlations of the image, which correspond to the features that allow to assess the quality of the generative network. It is important to note that a simple pixel-wise difference would not be able to account for the strong inter-dependencies of the pixels when they are spatially close.
The structural similarity index allows accounting for the similarities between images that just differ on the hidden variables, in particular focusing on
both luminance masking and contrast masking terms~\cite{Brunet2012} that appear due to the peaks of the dynamical correlators.

\subsection{Discriminator}
\begin{figure*}[t!]
    \includegraphics[width=.8\textwidth]{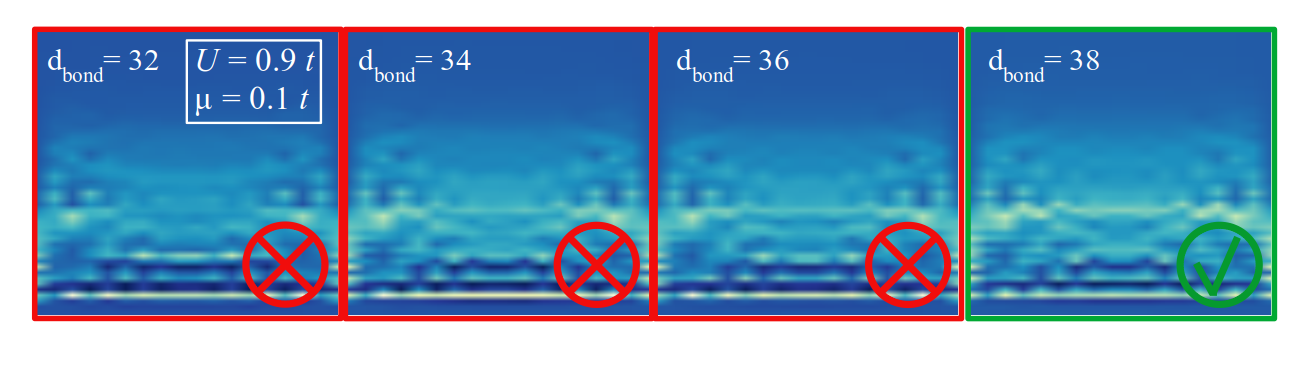}
    \caption{Outlier detection in the Hubbard model for calculations with bond
    dimensions $d_{bond}=30-40$. It is observed that, while the images show very strong similarities,
    the discriminator is capable of distinguishing fluctuations coming from ill-convergence and coming from the hidden variables.
    }
        \label{fig:subset_hubbard_disc_2}
\end{figure*}

The cGAN algorithm is trained with high-quality many-body data, with $d_{bond}$ large enough so that all the results have converged. The training examples contain however different values for the hidden variables, that give rise to small differences in the dynamical correlators. While both the hidden variables and under converge in $d_{bond}$ give rise to small differences, distinguishing them is a highly non-trivial task. Results with different hidden values are physically significant, whereas ill-converged results in $d_{bond}$ are outliers that should be flagged as defective data. The key point of our outlier detection is that our algorithm learns to distinguish fluctuations arising from the hidden variables from fluctuations coming from an ill-converged result.
While for a human is clearly easy to distinguish different images that have small fluctuations, distinguishing if the fluctuations correspond to different hidden variables (i.e. acceptable data) or ill-convergence (i.e. defective data) is a remarkably challenging problem even for human experts. To the best of our knowledge, no algorithm exists for performing this distinction. 

To demonstrate that the outlier detection also allows to flag unphysical results in the Hubbard model, we show in Fig.~\ref{fig:subset_hubbard_disc_2} the dynamical correlators for the Hubbard chain with higher bond dimensions, in particular from $d_{bond}=30-40$. It is clearly observed that, while the dynamical correlators show a high similarity displaying small fluctuations, the discriminator is capable of distinguishing the small fluctuations coming from an ill-converged calculations from the fluctuations of the hidden variables.

\bibliography{references}{}

\end{document}